\documentclass{elsart}
\usepackage{amsbsy,amssymb,amstext,graphicx}

\makeatletter
\def\eqsecnum{%
        \@addtoreset{equation}{section}%
        \def\theequation{\arabic{section}.\arabic{equation}}%
}
\eqsecnum
\newcommand{\be}{\begin{equation}}
\newcommand{\ee}{\end{equation}}
\newcommand{\ba}{\begin{array}}
\newcommand{\ea}{\end{array}}
\newcommand{\bea}{\begin{eqnarray}}
\newcommand{\eea}{\end{eqnarray}}
\newcommand{\bean}{\begin{eqnarray*}}
\newcommand{\eean}{\end{eqnarray*}}

\newcommand{\lp}{\left(}
\newcommand{\rp}{\right)}
\newcommand{\ls}{\left[}
\newcommand{\rs}{\right]}
\newcommand{\lc}{\left\{}
\newcommand{\rc}{\right\}}

\renewcommand{\d}{\mbox{\rm d}}
\newcommand{\g}{\text{\slshape g}}
\newcommand{\gb}%
	{\text{\bf\slshape g%
	}}
\newcommand{\p}{\partial}

\newcommand{\veps}{\varepsilon}
\newcommand{\vphi}{\varphi}

\newcommand{\kB}{k_{\rm B}}
\newcommand{\const}{\mathop{{\rm const}}\nolimits}
\newcommand{\ddt}{\frac{\partial}{\partial t}}
\newcommand{\dd}[1]{\frac{\partial}{\partial #1}}
\newcommand{\refp}[1]{(\ref{#1})}
\newcommand{\ds}{\displaystyle}
\newcommand{\scs}{\scriptstyle}

\newcommand*{\hs}{\hat{\sigma}}
\newcommand*{\thsgb}{\theta(\hat{\sigma}\gb)}
\newcommand*{\tmhsgb}{\theta(-\hat{\sigma}\gb)}
\newcommand*{\rt}{(\vec{r}_1;t)}
\newcommand*{\n}{\phantom{1}}
\newcommand*{\bi}{\bibitem}
\newcommand*{\vt}{\!\vartriangle\!\!}
\newcommand*{\vts}{\vartriangle}

\begin{document}
\begin{frontmatter}
\title{Kinetic equation for liquids with a multistep potential of
interaction. II: Calculation of transport coefficients}

\author{M.V.Tokarchuk, I.P.Omelyan, A.E.Kobryn}

\address{Institute for Condensed Matter Physics\\
	Ukrainian National Academy of Sciences,\\
	1~Svientsitskii St., UA--79011 Lviv, Ukraine}

\begin{abstract}
We consider a new kinetic equation for systems with a multistep potential of 
interaction proposed by us recently in Physica A 234 (1996) 89. This 
potential consists of the hard sphere part and a system of attractive and 
repulsive walls. Such a model is a generalization of many previous 
semi-phenomenological kinetic theories of dense gases and liquids. In this 
article a normal solution to the new kinetic equation has been obtained, 
integral conservation laws in the first order on gradients of hydrodynamic 
parameters have been derived as well. The expressions for transport 
coefficients are calculated for the case of stationary process. We also 
consider limiting cases for this kinetic equation. For specific parameters 
of model interaction potential in shape of the multistep function, the 
obtained results rearrange to those of previous kinetic theories by means of 
the standard Chapman-Enskog method. In view of this, new theory can be 
considered as a generalized one which in some specific cases arrives at the 
results of previous ones and in such a way displays the connection of these 
theories between themselves. At the end of this article we present results 
of numerical computation of transport coefficients for Argon along curve of 
saturation and their comparison with experimental data available and MD 
simulations.
\end{abstract}

\end{frontmatter}
\newpage

\setlength{\parskip}{3pt plus1pt minus2pt}

\section{Introduction}

The main problem of kinetic theory in the last few decades was to 
appreciate properties of dense gases and liquids on the basis of motion 
laws and molecular interactions. In 1872 L.Boltzmann suggested a kinetic 
equation for a dilute gas \cite{1}. At the same time, he constructed the 
entropy functional and proved $H$-theorem.

In 1946, N.N.Bogolubov suggested a new approach \cite{2} for the problem of 
derivation of kinetic equations, which describes irreversible processes in 
gases. He aspired to change suggestions of the Boltzmann's theory to a more 
rigorous ones and generalize the theory to higher densities. In this 
theory the Boltzmann kinetic equation is obtained in the zeroth 
approximation on density. It is possible at the description of gas kinetics 
on rough enough time scale and at the presumption that quantities, 
which define gas state, can be expanded into a series on the density. 
However, after studying of higher order expansion terms (starting from the 
second one) their divergency has been found \cite{3}. The reason for such a 
divergency lies in the fact that the accepted expansion is based on a 
dynamics of isolated groups of particles in an infinite space without 
taking into consideration the whole media, i.e. without taking into 
account all others particles. In contrast to an equilibrium liquid, 
velocity correlations in nonequilibrium state include long-range 
correlations between particles. It does not allow to generalize the 
Boltzmann kinetic equation to high densities.

It is particularly remarkable that a successful empirical kinetic theory of 
dense gases was developed by Enskog in 1922 at least for the case of hard 
spheres (standard Enskog theory -- SET) \cite{4}. His arguments were 
similar to the Boltzmann's ones. In 1961 Davis {\slshape et all} \cite{5} 
suggested the DRS kinetic theory, where the so-called ``square-well'' 
potential of interaction is considered. DRS theory is the analogue of the 
usual SET theory for a specific type of interaction potential. Here, an 
attractive part of real interaction potential is approximated by some 
finite attractive wall. A revised Enskog theory RET \cite{5d,6,7} and a 
revised version of DRS -- RDRS \cite{8} have been obtained. The necessity 
of the revised versions could be explained as follows. I: kinetic equations 
of initial versions of those theories cannot be derived sequentially. II: 
exact entropy functional cannot be constructed, as a result $H$-theorem 
cannot be proved. But some recipe of construction of entropy functional 
for the SET theory was suggested in papers \cite{5a,5b,5c} The exact 
entropy functional and proof of the $H$-theorem were constructed for 
revised versions RET \cite{6,7} and RDRS \cite{8} only. Furthermore, the 
RET kinetic equation has been obtained successfully within the frame of 
some theoretical scheme, which is analogous to Bogolubov one \cite{2}, but 
using a modification of boundary conditions. The last ones take into 
account local conservation laws in the solution of the BBGKY hierarchy 
\cite{9}.

For the purpose to use results of the SET theory to real systems, Enskog 
suggested to change hydrostatic pressure of a system of hard spheres to a 
thermodynamic pressure of a real system. Having this assertion in mind, 
Hanley {\slshape et all} \cite{10} built new kinetic theory MET (modified 
Enskog theory), where hard sphere diameter $\sigma$ is defined via the 
second virial coefficient of the equation of state of a system. In such a 
way, $\sigma$ becomes dependent on temperature and density. Using different 
equations of state: BH \cite{11}, WCA \cite{12}, MC/RS \cite{13,14} and 
others, one obtains corresponding versions of MET.

In the kinetic mean field theory (KMFT) \cite{15} next to the hard sphere 
interaction potential one considers also some smooth attractive ``tail''. 
It is noted in \cite{16} that in this case the quasiequilibrium binary 
correlation function of hard spheres should be replaced to a such one, 
which takes into account explicitly or implicitly the total interaction 
potential. The main conclusion of KMFT is that the smooth part of 
interaction potential in the first approximation on gradients of 
hydrodynamic parameters does not contribute explicitly into transport 
coefficients. There is only an indirect contribution via the binary 
correlation function. Its dependency on temperature is defined by a 
smooth part of the interaction potential.

In our recent paper \cite{17} we suggested a new kinetic equation for 
systems with a multistep potential of interaction (MSPI). This potential 
consists of the hard sphere part and a system of attractive and repulsive 
walls. Such a model is a generalization of SET (RET, MET), DRS (RDRS) and 
KMFT theories. We also proved for this equation $H$-theorem. But normal 
solution was not published yet. In this article going to the schema of 
construction of normal solutions of kinetic equations with the help of 
boundary conditions method \cite{18}, a normal solution to the new kinetic 
equation has been obtained, integral conservation laws in the first order 
on gradients of hydrodynamic parameters have been derived as well. The 
expressions for such transport coefficients as bulk and shear viscosity and 
thermal conductivity are calculated for the case of stationary process. We 
also consider limiting cases for this kinetic equation. For specific 
parameters of model interaction potential in shape of the multistep 
function, the obtained results rearrange to those of the SET (RET, MET), 
DRS (RDRS) or KMFT theories by means of the standard Chapman-Enskog method 
\cite{19}. In view of this, new theory can be considered as a generalized 
one which in some specific cases arrives at the results of previous ones 
and in such a way displays the connection of these theories between 
themselves. In section 9 we present results of numerical computation of 
transport coefficients for Argon and their comparison with experimental 
data available and MD simulations.

\section{Multistep potential of interaction. Kinetic equation}

\begin{figure}[ht]
\begin{center}
\fbox{%
\includegraphics*[bb=177 305 421 537]{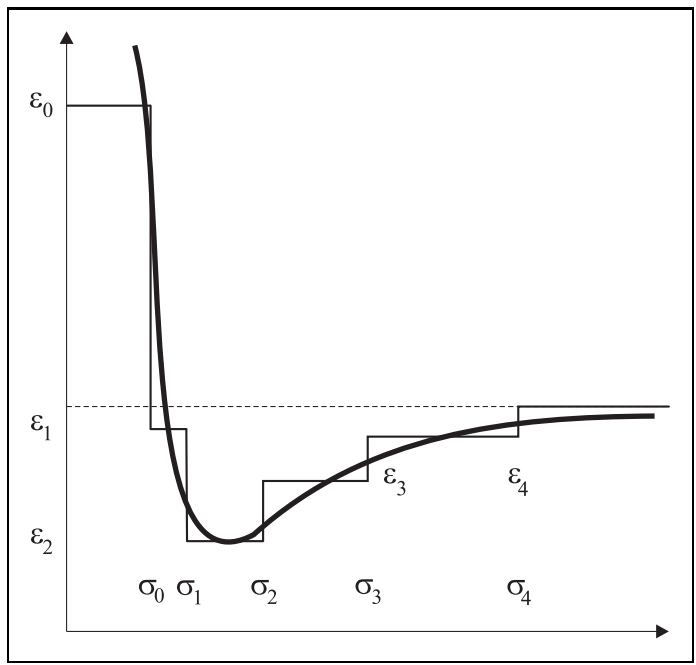}}
\end{center}
\caption{Multistep potential of interactin with parameters $n^*=1$, 
$m^*=3$.}
\label{figure1}
\end{figure}

Let us consider a system of $N$ classical particles in volume $V$ when 
$N\to\infty$ and $V\to\infty$ provided $N/V=\const$. The Hamiltonian of 
this system reads:
%
%
\bea
H_N=\sum_{i=1}^N\frac{p^2}{2m}+\sum_{i<j}\varphi_{ij}.\label{e2.1}
\eea
Our purpose is the most detailed analysis of nonequilibrium processes in 
dense systems. To do this let us define MSPI 
$\vphi_{ij}\equiv\vphi(|\vec{r}_i-\vec{r}_j|)\equiv\vphi(|\vec{r}_{ij}|)
\equiv\vphi(r_{ij})$ in a form of the multistep function:
%
%
\bea
\vphi_{ij}=\left\{
\begin{array}{cllccll}
\infty&;\quad&&&r_{ij}&<&\sigma_0,\\
\varepsilon_k&;&\sigma_{k-1}&<&r_{ij}&<&\sigma_k;\quad k=1,\ldots,N^*,\\
0&;&\sigma_{N^*}&<&r_{ij}.
\end{array}\right.
\label{e2.2}
\eea
Here $N^*$ is the total number of attractive and repulsive walls except 
hard sphere one. For our convenience we distinguish separately systems 
of attractive and repulsive walls. Let one has $n^*$ attractive walls, they 
are separated by the distances $\sigma_{li}$ and have heights 
$\vt\veps_{li}$, $i=1,\ldots,n^*$; and $m^*$ repulsive walls with 
parameters $\sigma_{rj}$ and $\vt\veps_{rj}$, $j=1,\ldots,m^*$, 
correspondingly. $\sigma_0$ is the location of hard sphere wall. It is 
obviously that $n^*+m^*=N^*$, $\vt\veps_{li}=\veps_{li}-\veps_{li+1}$, 
$\vt\veps_{rj}=\veps_{rj+1}-\veps_{rj}$. In such a way, parameters 
$\sigma_0$, $n^*$, $\sigma_{li}$, $\vt\veps_{li}$, $m^*$, $\sigma_{rj}$, 
$\vt\veps_{rj}$ define multistep potential of interaction completely 
(see Fig.~\ref{figure1}).

\begin{figure}[ht]
\begin{centering}
\setlength{\unitlength}{1cm}
\begin{picture}(11,8)
\put(0.5,1){%
\begin{picture}(10,10)
\put(0,3){\line(0,1){3}}
\put(1.5,5.5){\vector(-1,0){1.5}}
\put(0,5){\vector(1,0){1.5}}
\put(1.6,5){\makebox(0,0)[lc]{``$a$''}}
\put(0,2.75){\makebox(0,0)[l]{$\sigma_0$}}
\put(1.5,1.5){\vector(0,-1){1.5}}
\put(1.5,1.5){\vector(0,1){1.5}}
\put(1.4,1.5){\makebox(0,0)[rc]{$\vartriangle\!\!\varepsilon_{li}$}}
\put(3,0){\line(0,1){3}}
\put(2,2.5){\vector(1,0){1}}
\put(3,2){\vector(-1,0){1}}
\put(3,2.5){\vector(1,0){1}}
\put(4,2){\vector(-1,0){1}}
\put(4,1){\vector(-1,0){1}}
\put(3,0.5){\vector(1,0){1}}
\put(4.1,2.5){\makebox(0,0)[lc]{``$b_l$''}}
\put(4.1,2.0){\makebox(0,0)[lc]{``$c_l$''}}
\put(4.1,1.0){\makebox(0,0)[lc]{``$d_l$''}}
\put(3,-0.25){\makebox(0,0)[cc]{$i$, $\sigma_{li}$}}
\put(7,0.25){\line(0,1){2.5}}
\put(7,2.5){\vector(-1,0){1}}
\put(6,2){\vector(1,0){1}}
\put(8,2.5){\vector(-1,0){1}}
\put(7,2){\vector(1,0){1}}
\put(6,1){\vector(1,0){1}}
\put(7,0.5){\vector(-1,0){1}}
\put(5.9,2.5){\makebox(0,0)[rc]{``$b_r$''}}
\put(5.9,2.0){\makebox(0,0)[rc]{``$c_r$''}}
\put(5.9,1.0){\makebox(0,0)[rc]{``$d_r$''}}
\put(7,-0.25){\makebox(0,0)[cc]{$j$, $\sigma_{rj}$}}
\put(8.5,1.5){\vector(0,-1){1.25}}
\put(8.5,1.5){\vector(0,1){1.25}}
\put(8.6,1.5){\makebox(0,0)[lc]{$\vartriangle\!\!\varepsilon_{rj}$}}
\multiput(0,3)(0.4,0){22}{\line(1,0){0.2}}
\multiput(3,0)(0.4,0){6}{\line(1,0){0.2}}
\multiput(3,0)(-0.4,0){5}{\line(1,0){0.2}}
\multiput(7,0.25)(-0.4,0){6}{\line(-1,0){0.2}}
\multiput(7,0.25)(0.4,0){4}{\line(1,0){0.2}}
\multiput(7,2.75)(0.4,0){4}{\line(1,0){0.2}}
\put(-0.5,-1){\framebox(11,8){}}
\end{picture}}
\end{picture}\\
\end{centering}
\caption{Types of possible interactions (schematic draw).}
\label{figure2}
\end{figure}
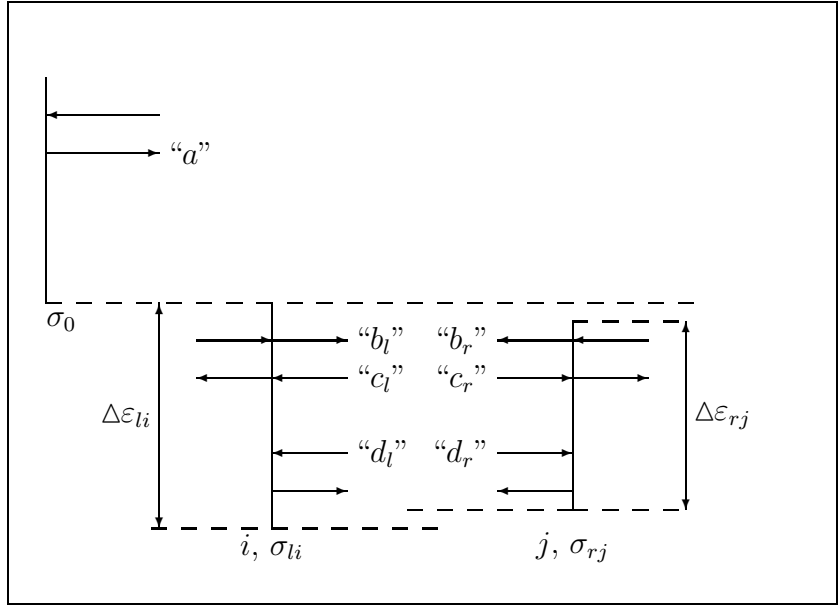

Going similarly to the derivation of the kinetic equation of the RET theory 
\cite{6,7,20} and taking into account the system of attractive and 
repulsive walls, one obtains the following kinetic equation:
%
%
\bea
\lp\ddt+\vec{v}_1\dd{\vec{r}_1}\rp f_1(x_1;t)&=&\int\d x_2\;\hat{T}
f_2(x_1,x_2;t),\label{e2.3}\\
f_2(x_1,x_2;t)&=&\g_2^{\mathrm{q}}\Big(\vec{r}_1,\vec{r}_2|n(t),\beta(t)\Big)
f_1(x_1;t)f_1(x_2;t),\nonumber
\eea
where $\g_2^{\mathrm{q}}\Big(\vec{r}_1,\vec{r}_2|n(t),\beta(t)\Big)$ is 
defined in the usual way from maximum of the entropy functional and in its 
turn is the functional of local values of density $n\rt$ and temperature 
$\beta\rt$. $\hat{T}$ is an operator which describes interaction of two 
particles at presence of MSPI:
%
%
\bea
\hat{T}&=&\hat{T}_{\mathrm{hs}}^a+\sum_{i=1}^{n^*}\hat{T}_{li}+
\sum_{j=1}^{m^*}\hat{T}_{rj},\label{e2.4}\\
\hat{T}_{\mathrm{hs}}^a&=&\sigma_0^2\int\d\hat{\sigma}\;
\hs\gb\;\theta(\hat{\sigma}\gb)
\lc\delta\lp\vec{r}_1-\vec{r}_2+\hat{\sigma}\sigma_0^+\rp
B_a(\hat{\sigma})-\delta\lp\vec{r}_1-\vec{r}_2-\hat{\sigma}\sigma_0^+\rp\rc.
\nonumber\\ \label{e2.5}
\eea
The last expression is nothing but operator of hard spheres interaction 
\cite{21}, $\hs$ is the unit vector along distance from the second particle 
to the first one, $\gb=\vec{v}_2-\vec{v}_1$ is relative velocity. 
$B_a(\hs)$ is the velocities shift operator as in classical mechanics of
elastic collisions. $\hat{T}_{li}=\sum_{k=b,c,d}\hat{T}_{lik}$ is an 
interaction operator on $i$th repulsive wall; 
$\hat{T}_{rj}=\sum_{k=b,c,d}\hat{T}_{rjk}$ is an interaction operator on 
$j$th attractive wall; $a$, $b$, $c$ are types of possible interactions 
(see Fig.~\ref{figure2}).
%
%
\bea
&&\hat{T}_{{li\atop rj}k}=\sigma_{{li\atop rj}}^2\int\d\hat{\sigma}\;
\hs\gb\;\theta_{{li\atop rj}k}(\ldots)\times{}\label{e2.6}\\
&&\lc\delta\lp\vec{r}_1-\vec{r}_2+\hat{\sigma}\sigma_{{li\atop rj}}^+\rp
B_{{li\atop rj}k}(\hat{\sigma})-
\delta\lp\vec{r}_1-\vec{r}_2-\hat{\sigma}\sigma_{{li\atop rj}}^+\rp\rc,
\nonumber
\eea
where
%
%
\bea
\ba{lclclcl}
\theta_{lib}&\equiv&\theta(-\hat{\sigma}\gb),&\quad&
\theta_{rjb}&\equiv&\theta(-\hat{\sigma}\gb),\\
\theta_{lic}&\equiv&\ds\theta\lp\hat{\sigma}\gb-
\frac{4\vt\veps_{li}}{m}\rp,
&\quad&\theta_{rjc}&\equiv&\ds\theta\lp-\hat{\sigma}\gb-
\frac{4\vt\veps_{rj}}{m}\rp,\\
\theta_{lid}&\equiv&
\ds\theta\lp\frac{4\vt\veps_{li}}{m}-\hat{\sigma}\gb\rp
\theta(\hat{\sigma}\gb),&\quad&
\theta_{rjd}&\equiv&
\ds\theta\lp\frac{4\vt\veps_{li}}{m}+\hat{\sigma}\gb\rp
\theta(-\hat{\sigma}\gb).
\ea\label{e2.7}
\eea
Expressions \refp{e2.7} are nothing but conditions for specific type of 
interaction ($\theta(z)$ is the Heaviside unit step function).

$B_{{li\atop rj}k}(\hs)$ is an operator which acts on velocities and 
changes them in accordance with the interaction of specific type on each 
wall:
%
%
\bea
B_{{li\atop rj}k}(\hat{\sigma})\Psi(\vec{v}_1,\vec{v}_2)=
\Psi(\vec{v}_{1{l\atop r}k},\vec{v}_{2{l\atop r}k}).\label{e2.8}
\eea
Type ``$a$'':
%
%
\bea
\ba{ll}
\vec{v}_1'&=\vec{v}_1+\hat{\sigma}(\hat{\sigma}\gb);\\
\vec{v}_2'&=\vec{v}_1-\hat{\sigma}(\hat{\sigma}\gb);\\
\ea\label{e2.9}
\eea
type ``$b$'':
%
%
\bea
\ba{ll}
\vec{v}_{1l}''&\ds=\vec{v}_1+\frac12\hs\lp\hs\gb+\sqrt{(\hs\gb)^2+
\frac{4\vt\veps_{li}}{m}}\rp;\\ [2ex]
\vec{v}_{1r}''&\ds=\vec{v}_1+\frac12\hs\lp\hs\gb-\sqrt{(\hs\gb)^2+
\frac{4\vt\veps_{rj}}{m}}\rp;\\
\ea\label{e2.10}
\eea
type ``$c$'':
%
%
\bea
\ba{ll}
\vec{v}_{1l}'''&\ds=\vec{v}_1+\frac12\hs\lp\hs\gb-\sqrt{(\hs\gb)^2-
\frac{4\vt\veps_{li}}{m}}\rp;\\ [2ex]
\vec{v}_{1r}'''&\ds=\vec{v}_1+\frac12\hs\lp\hs\gb+\sqrt{(\hs\gb)^2-
\frac{4\vt\veps_{rj}}{m}}\rp;\\
\ea\label{e2.11}
\eea
type ``$d$'':
%
%
\bea
\vec{v}_{1{l\atop r}}'''=\vec{v}_1+\hat{\sigma}(\hat{\sigma}\gb)
\equiv\vec{v}_1'.\label{e2.12}
\eea
Now kinetic equation \refp{e2.3} can be rewritten in an explicit form:
%
%
\bea
&&\text{\small$\ds\lp\ddt+\vec{v}_1\dd{\vec{r}_1}\rp f_1(x_1;t)={}$}
\label{e2.13}\\
&&\text{\small$\ds\sigma_0\int\d\hs\d\vec{v}_2\;\hs\gb\;\thsgb
\lc f_2(\vec{r}_1,\vec{v}_1',\vec{r}_1+\hs\sigma_0^+,\vec{v}_2';t)-
f_2(x_1,\vec{r}_1-\hs\sigma_0^+,\vec{v}_2;t)\rc+{}$}\nonumber\\
&&\text{\small$\ds\sum_{i=1}^{n^*}\sigma_{li}^2\int\d\hs\d\vec{v}_2\;
\hs\gb\times{}$}\nonumber\\
&&\text{\small$\ds\ls\tmhsgb
\lc f_2(\vec{r}_1,\vec{v}_{1l}'',
\vec{r}_1+\hs\sigma_{li}^+,\vec{v}_{2l}'';t)-
f_2(x_1,\vec{r}_1-\hs\sigma_{li}^-,\vec{v}_2;t)\rc\right.+{}$}\nonumber\\
&&\text{\small$\ds\theta\lp\hs\gb-{\ds\sqrt{\frac{4\vt\veps_{li}}{m}}}\rp
\lc f_2(\vec{r}_1,\vec{v}_{1l}''',
\vec{r}_1+\hs\sigma_{li}^-,\vec{v}_{2l}''';t)-
f_2(x_1,\vec{r}_1-\hs\sigma_{li}^+,\vec{v}_2;t)\rc+{}$}\nonumber\\
&&\text{\small$\ds\thsgb\theta
\lp{\ds\sqrt{\frac{4\vt\veps_{li}}{m}}}-\hs\gb\rp\times{}$}\nonumber\\
&&\text{\small$\ds\left.\lc f_2(\vec{r}_1,\vec{v}_{1l}'''',
\vec{r}_1+\hs\sigma_{li}^+,\vec{v}_{2l}'''';t)-
f_2(x_1,\vec{r}_1-\hs\sigma_{li}^+,\vec{v}_2;t)\rc\rs+{}$}\nonumber\\
&&\text{\small$\ds\sum_{j=1}^{m^*}\sigma_{rj}^2\int\d\hs\d\vec{v}_2\;
\hs\gb\times{}$}\nonumber\\
&&\text{\small$\ds\ls\thsgb
\lc f_2(\vec{r}_1,\vec{v}_{1r}'',
\vec{r}_1+\hs\sigma_{rj}^-,\vec{v}_{2r}'';t)-
f_2(x_1,\vec{r}_1-\hs\sigma_{rj}^+,\vec{v}_2;t)\rc\right.+{}$}\nonumber\\
&&\text{\small$\ds\theta\lp-\hs\gb-{\ds\sqrt{\frac{4\vt\veps_{rj}}{m}}}\rp
\times{}$}\nonumber\\
&&\text{\small$\ds\lc f_2(\vec{r}_1,\vec{v}_{1r}''',
\vec{r}_1+\hs\sigma_{rj}^+,\vec{v}_{2r}''';t)-
f_2(x_1,\vec{r}_1-\hs\sigma_{rj}^-,\vec{v}_2;t)\rc+{}$}\nonumber\\
&&\text{\small$\ds\tmhsgb\theta
\lp{\ds\sqrt{\frac{4\vt\veps_{rj}}{m}}}+\hs\gb\rp\times{}$}\nonumber\\
&&\text{\small$\ds\left.\lc f_2(\vec{r}_1,\vec{v}_{1r}'''',
\vec{r}_1+\hs\sigma_{rj}^-,\vec{v}_{2r}'''';t)-
f_2(x_1,\vec{r}_1-\hs\sigma_{rj}^-,\vec{v}_2;t)\rc\rs={}$}\nonumber\\
&&\text{\small$\ds J_{\mathrm{E}}(f_1,f_1)+\sum_{i=1}^{n^*}
\lp J_{lib}+J_{lic}+J_{lid}\rp+\sum_{j=1}^{m^*}
\lp J_{rjb}+J_{rjc}+J_{rjd}\rp$.}\nonumber
\eea
One can draw the following conclusions regarding the form of \refp{e2.13}. 
In the absence of attractive and repulsive walls ($\vt\veps_{li}=0$, 
$i=1,\ldots,n^*$, $\vt\veps_{rj}=0$, $j=1,\ldots,m^*$) this kinetic 
equation transfers to that one of the RET theory \cite{6,7}. In the 
presence of only one finite attractive wall ($\vt\veps_{li}=0$, 
$i=1,\ldots,n^*$, $\vt\veps_{rj}=0$, $j=2,\ldots,m^*$, $\vt\veps_{r1}\ne0$) 
one obtains the kinetic equation of the RDRS theory \cite{8}. Moreover, it 
can be shown, in the third special case when the set of walls is merged 
with some smooth potential $\phi_t$ and
$\vt\sigma_{li}=\sigma_{li}-\sigma_{li-1}\to0$, $i=1,\ldots,n^*-1$, 
$\vt\veps_{li}\to0$, $n^*\to\infty$, 
$\vt\sigma_{rj}=\sigma_{rj+1}-\sigma_{rj}\to0$, $j=1,\ldots,m^*-1$, 
$\vt\veps_{rj}\to0$, $m^*\to\infty$, and
\bean
-\frac{\vt\veps_{li}}{\vt\sigma_{li}}&\to&\phi'_t(\sigma_{li}),\\
\frac{\vt\veps_{rj}}{\vt\sigma_{rj}}&\to&\phi'_t(\sigma_{rj}),\\
\eean
kinetic equation \refp{e2.13} transfers to those one of the KMFT theory 
\cite{15}.

\section{Macroscopic conservation laws. The zeroth approximation}

Let us introduce the following set of hydrodynamic parameters: particles 
number density $n(\vec{r};t)$, hydrodynamic velocity $\vec{u}(\vec{r};t)$, 
densities of kinetic $\omega_{\mathrm{k}}$ and interaction 
$\omega_{\mathrm{i}}$ energies:
%
%
\bea
\ba{rcl}
n\rt&=&\ds\int\d\vec{v}_2\;f_1(x_1;t),\\
\vec{u}\rt&=&\ds\int\d\vec{v}_2\;f_1(x_1;t)\frac{\vec{v}_1}{n\rt},\\ [1ex]
\omega_{\mathrm{k}}\rt&=&\ds\int\d\vec{v}_2\;f_1(x_1;t)
\frac{c_1^2}{2n\rt},\\
\omega_{\mathrm{i}}\rt&=&\ds\int\d\vec{r}_2\;n(\vec{r}_2;t)
\g_2^{\mathrm{q}}(\vec{r}_1,\vec{r}_2|n(t),\beta(t))
\frac{1}{2m}\Phi(|\vec{r}_{12}|).\\
\ea\label{e3.1}
\eea
We also introduce the set of interaction invariants 
$\vec{\Psi}$$\equiv$$\lc\! m,m\vec{v},\frac12\!\ls mv^2\mbox{+}
\phi(r_{12})\rs\!\rc$. Multiplying initial kinetic equation \refp{e2.13} on 
each component of the vector $\vec{\Psi}$ and integrating with respect to 
$\vec{v}_2$ one obtains the equation of continuity, the equation of motion 
and equation of balance of kinetic energy, correspondingly:
%
%
\bea
\ba{rcl}
\ds\frac{1}{n}\frac{\d n}{\d t}&=&\ds-\dd{r_{1\alpha}}u_{\alpha},\\
\ds\frac{\d u_\alpha}{\d t}&=&
\ds-\frac{1}{mn}\dd{r_{1\beta}}P_{\alpha\beta},\\
\ds\frac{\d\omega_{\mathrm{k}}}{\d t}&=&
\ds-\frac{1}{mn}\lc\dd{r_{1\alpha}}q_\alpha+
P_{\alpha\beta}\dd{r_{1\beta}}u_\alpha\rc,\\
\ea
\quad\frac{\d}{\d t}=\ddt+u_\gamma\dd{r_{1\gamma}}.\label{e3.2}
\eea
The equation for interaction energy density one obtains automatically 
differentiating the expression for $\omega_{\mathrm{i}}$ taking into 
account conservation laws for $n(\vec{r};t)$, $\vec{u}(\vec{r};t)$ and 
$\omega_{\mathrm{k}}$ \refp{e3.2}. Stress tensor $P_{\alpha\beta}$ and heat 
flow vector consist on kinetic and interaction parts:
%
%
\bea
\ba{lclcll}
P_{\alpha\beta}&=&P_{\alpha\beta}^{\mathrm{k}}&+&
P_{\alpha\beta}^{\mathrm{i}}&,\\
q_\alpha&=&q_\alpha^{\mathrm{k}}&+&q_\alpha^{\mathrm{i}}&.\\
\ea\label{e3.3}
\eea
In its turn, the potential parts consist of the hard sphere term and 
components conditioned by the set of attractive and repulsive walls:
%
%
\bea
\ba{llllllllllll}
P_{\alpha\beta}^{\mathrm{i}}&=P_{1\alpha\beta}^{\mathrm{i}}&+
\!\ds\sum_{i=1}^{n^*}(P_{i2\alpha\beta}^{\mathrm{i}}&+&
P_{i3\alpha\beta}^{\mathrm{i}}&+&P_{i4\alpha\beta}^{\mathrm{i}})&+
\ds\sum_{j=1}^{m^*}(P_{j2\alpha\beta}^{\mathrm{i}}&+&
P_{j3\alpha\beta}^{\mathrm{i}}&+&P_{j4\alpha\beta}^{\mathrm{i}}),\\
q_\alpha^{\mathrm{i}}&=q_{1\alpha}^{\mathrm{i}}&+
\!\ds\sum_{i=1}^{n^*}(q_{i2\alpha}^{\mathrm{i}}&+&
q_{i3\alpha}^{\mathrm{i}}&+&q_{i4\alpha}^{\mathrm{i}})&+
\ds\sum_{j=1}^{m^*}(q_{j2\alpha}^{\mathrm{i}}&+&
q_{j3\alpha}^{\mathrm{i}}&+&q_{j4\alpha}^{\mathrm{i}}),\\
\ea\nonumber\\\label{e3.4}
\eea
where indices 1, 2, 3, and 4 correspond to different types of interactions: 
``$a$'', ``$b$'', ``$c$'', and ``$d$'', correspondingly. Now let us write 
down the expressions for stress tensor $P_{\alpha\beta}$ and heat flow 
vector $q_\alpha$ in an explicit form:
%
%
\bea
\ba{ll}
P_{\alpha\beta}^{\mathrm{k}}&=\ds\int\d\vec{v}_1\;f_1(x_1;t)
mc_{1\alpha}c_{1\beta},\\
q_\alpha^{\mathrm{k}}&=\ds\int\d\vec{v}_1\;f_1(x_1;t)
\frac{mc_1^2}{2}c_{1\alpha},\\
\ea\label{e3.5}
\eea
%
%
\bea
P_{1\alpha\beta}^{\mathrm{i}}&=&\frac12m\sigma_0^3\int\d\vec{v}_1\d\vec{v}_2
\d\hs\;\hs\gb\;\thsgb\lp v_{1\alpha}'-v_{1\alpha}\rp\hs_\beta\times{}
\label{e3.6}\\
&&\int_0^1\d\lambda\;f_2(\vec{r}_1+\lambda\hs\sigma_0^+,\vec{v}_1,
\vec{r}_1+\lambda\hs\sigma_0^+-\hs\sigma_0^+,\vec{v}_2;t),\nonumber\\
q_{1\alpha}^{\mathrm{i}}&=&\frac12m\sigma_0^3\int\d\vec{v}_1\d\vec{v}_2
\d\hs\;\hs\gb\;\thsgb\lp\frac{\lp c_1'\rp^2}{2}-
\frac{c_1^2}{2}\rp\hs_\alpha\times{}\label{e3.7}\\
&&\int_0^1\d\lambda\;f_2(\vec{r}_1+\lambda\hs\sigma_0^+,\vec{v}_1,
\vec{r}_1+\lambda\hs\sigma_0^+-\hs\sigma_0^+,\vec{v}_2;t),\nonumber\\
P_{i2\alpha\beta}^{\mathrm{i}}&=&\frac12m\sigma_{li}^3
\int\d\vec{v}_1\d\vec{v}_2\d\hs\;\hs\gb\;\tmhsgb
\lp v_{1l\alpha}''-v_{1\alpha}\rp\hs_\beta\times{}\label{e3.8}\\
&&\int_0^1\d\lambda\;f_2(\vec{r}_1+\lambda\hs\sigma_{li}^-,\vec{v}_1,
\vec{r}_1+\lambda\hs\sigma_{li}^--\hs\sigma_{li}^-,\vec{v}_2;t),\nonumber\\
q_{i2\alpha}^{\mathrm{i}}&=&\frac12m\sigma_{li}^3
\int\d\vec{v}_1\d\vec{v}_2\d\hs\;\hs\gb\;\tmhsgb
\lp\frac{\lp c_{1l}''\rp^2}{2}-\frac{c_1^2}{2}-\frac{\vt\veps_{li}}{m}\rp
\hs_\alpha\times{}\label{e3.9}\\
&&\int_0^1\d\lambda\;f_2(\vec{r}_1+\lambda\hs\sigma_{li}^-,\vec{v}_1,
\vec{r}_1+\lambda\hs\sigma_{li}^--\hs\sigma_{li}^-,\vec{v}_2;t),\nonumber\\
P_{i3\alpha\beta}^{\mathrm{i}}&=&\frac12m\sigma_{li}^3
\int\d\vec{v}_1\d\vec{v}_2\d\hs\;\hs\gb\;\theta\lp\hs\gb-
\sqrt{\frac{4\vt\veps_{li}}{m}}\rp
\lp v_{1l\alpha}'''-v_{1\alpha}\rp\hs_\beta\times{}\label{e3.10}\\
&&\int_0^1\d\lambda\;f_2(\vec{r}_1+\lambda\hs\sigma_{li}^+,\vec{v}_1,
\vec{r}_1+\lambda\hs\sigma_{li}^+-\hs\sigma_{li}^+,\vec{v}_2;t),\nonumber\\
q_{i3\alpha}^{\mathrm{i}}&=&\frac12m\sigma_{li}^3
\int\d\vec{v}_1\d\vec{v}_2\d\hs\;\hs\gb\;\theta\lp\hs\gb-
\sqrt{\frac{4\vt\veps_{li}}{m}}\rp
\ls\frac{\lp c_{1l}'''\rp^2}{2}-\frac{c_1^2}{2}+
\frac{\vt\veps_{li}}{2m}\rs\hs_\alpha\nonumber\\
&\times&\int_0^1\d\lambda\;f_2(\vec{r}_1+\lambda\hs\sigma_{li}^+,\vec{v}_1,
\vec{r}_1+\lambda\hs\sigma_{li}^+-\hs\sigma_{li}^+,\vec{v}_2;t),
\label{e3.11}\\
P_{i4\alpha\beta}^{\mathrm{i}}&=&\frac12m\sigma_{li}^3
\int\d\vec{v}_1\d\vec{v}_2\d\hs\;\hs\gb\;\thsgb\theta\lp
\sqrt{\frac{4\vt\veps_{li}}{m}}-\hs\gb\rp
\lp v_{1l\alpha}''''-v_{1\alpha}\rp\hs_\beta\times{}\nonumber\\
&&\int_0^1\d\lambda\;f_2(\vec{r}_1+\lambda\hs\sigma_{li}^+,\vec{v}_1,
\vec{r}_1+\lambda\hs\sigma_{li}^+-\hs\sigma_{li}^+,\vec{v}_2;t),
\label{e3.12}\\
q_{i4\alpha}^{\mathrm{i}}&=&\frac12m\sigma_{li}^3
\int\d\vec{v}_1\d\vec{v}_2\d\hs\;\hs\gb\;\thsgb\theta\lp
\sqrt{\frac{4\vt\veps_{li}}{m}}-\hs\gb\rp
\lp\frac{\lp c_{1l}''''\rp^2}{2}-\frac{c_1^2}{2}\rp\times{}\nonumber\\
&&\int_0^1\d\lambda\;f_2(\vec{r}_1+\lambda\hs\sigma_{li}^+,\vec{v}_1,
\vec{r}_1+\lambda\hs\sigma_{li}^+-\hs\sigma_{li}^+,\vec{v}_2;t).
\label{e3.13}
\eea
Expressions for $P_{j2\alpha\beta}$, $P_{j3\alpha\beta}$, 
$P_{j4\alpha\beta}$, $q_{j2\alpha}$, $q_{j3\alpha}$, and $q_{j4\alpha}$ 
look similar to \refp{e3.6}--\refp{e3.13} $P_{i2\alpha\beta}$, 
$P_{i3\alpha\beta}$, $P_{i4\alpha\beta}$, $q_{i2\alpha}$, $q_{i3\alpha}$, 
and $q_{ij4\alpha}$, correspondingly, at formal replacing 
$\sigma_{li}^{\pm}\to\sigma_{rj}^{\mp}$, $\hs\gb\to-\hs\gb$, 
$\vec{v}_{1l}\to\vec{v}_{1r}$, $\vt\veps_{li}\to\;\vt\veps_{rj}$.

At the end of this section we consider macroscopic conservation laws in the 
zeroth approximation. Let us suppose that one-particle distribution 
function in this case is equal to the local-equilibrium Maxwell one:
%
%
\bea
f_1\equiv f_1^{(0)}(x_1;t)=n\rt
\lp\frac{m}{2\pi\kB T\rt}\rp^{\scs3/2}\!\!\!\exp
\lc-\frac{mc_1^2\rt}{2\kB T\rt}\rc.\label{e3.14}
\eea
Neglecting by any spatial gradients one obtains
%
%
\bea
&&\g_2^{\mathrm{q}}(\vec{r}_1,\vec{r}_2|n\rt,\beta\rt)\simeq
\g_2^{\mathrm{eq}}\lp r_{12};n\lp\frac{\vec{r}_1+\vec{r}_2}{2};t\rp,
\beta\lp\frac{\vec{r}_1+\vec{r}_2}{2};t\rp\rp,\nonumber\\
\label{e3.15}\\
&&\int_0^1\d\lambda\;f_2(x_1,x_2;t)\simeq\ds\int_0^1\d\lambda\;f_2^{(0)}
(\ldots)=\g_2^{\mathrm{eq}}(\sigma|n,\beta)f_1^{(0)}(x_1;t)
f_1^{(0)}(\vec{r}_1,\vec{v}_2;t).\nonumber
\eea

It is well known from the equilibrium statistical mechanics \cite{22} that 
any sharp jump of interaction potential results in a corresponding jump of 
binary correlation function:
%
%
\bea
\frac{\g_2^{\mathrm{eq}}(\sigma_{li}^+|n,\beta)}
{\g_2^{\mathrm{eq}}(\sigma_{li}^-|n,\beta)}&=&
\exp\lc\beta\vt\veps_{li}\rc,\nonumber\\
\label{e3.16}\\
\frac{\g_2^{\mathrm{eq}}(\sigma_{rj}^-|n,\beta)}
{\g_2^{\mathrm{eq}}(\sigma_{rj}^+|n,\beta)}&=&
\exp\lc\beta\vt\veps_{rj}\rc.\nonumber
\eea
Substituting these relations into \refp{e3.5}--\refp{e3.13}, one obtains: 
$P_{\alpha\beta}=p\delta_{\alpha\beta}$, where $p$ is the hydrostatic 
pressure which consists on kinetic $p^{\mathrm{k}}$ and interaction 
$p^{\mathrm{i}}$ parts; $q_\alpha=0$:
%
%
\bea
\ba{lcl}
p&=&p^{\mathrm{k}}+p^{\mathrm{i}},\\
p^{\mathrm{k}}&=&n\kB T,\\
p^{\mathrm{i}}&=&\ds\frac{2}{3}\pi n^2\kB T\Lambda,\\
\Lambda&=&\text{\small$\ds\sigma_0^3\g_2(\sigma_0^+)-\sum_{i=1}^{n^*}
\sigma_{li}^3\g_2(\sigma_{li}^+)\lc\e^{-\beta\vts\veps_{li}}-1\rc
+\ds\sum_{j=1}^{m^*}
\sigma_{rj}^3\g_2(\sigma_{rj}^-)\lc\e^{-\beta\vts\veps_{rj}}-1\rc$}\!,\\
\g_2^{\mathrm{eq}}&\equiv&\g_2^{\mathrm{eq}}\Big(\sigma|n\rt,\beta\rt\Big).\\
\ea\label{3.17}
\eea
Then, starting from \refp{e3.2}, one has the conservation laws in the 
zeroth approximation (Euler laws):
%
%
\bea
\ba{rcl}
\ds\frac{1}{n}\frac{\d n}{\d t}&=&\ds-\frac{\p\vec{u}}{\p\vec{r}_1},\\ [1ex]
\ds\frac{\d\vec{u}}{\d t}&=&
\ds-\frac{1}{mn}\frac{\p P}{\p\vec{r}_1},\\ [1ex]
\ds\frac{\d\omega_{\mathrm{k}}}{\d t}&=&
\ds-\frac{1}{mn}p\frac{\p\vec{u}}{\p\vec{r}_1},
\quad\omega_{\mathrm{k}}=\frac{3}{2m\beta},\\ [1ex]
\omega_{\mathrm{i}}^{(0)}\rt&=&\ds\frac{1}{2m}\int\d\vec{r}_{12}\;
n\rt\g_2^{\mathrm{eq}}
\Big(|\vec{r}_{12}|;n\rt,\beta\rt\Big)\Phi(|\vec{r}_{12}|).\\
\ea\label{e3.18}
\eea

\section{The principle of construction of higher order approximations.
Normal solutions by means of boundary conditions method}

Let us consider initial kinetic equation \refp{e2.3} (or \refp{e2.13})
%
%
\bea
\lp\ddt+\vec{v}_1\dd{\vec{r}_1}\rp f_1(x_1;t)=
\int\d x_2\;\hat{T}f_2(x_1,x_2;t)=J(f_1,f_1),\label{e4.1}
\eea
where the collision integral $J(f_1,f_1)$ consists of the usual one of the 
RET theory -- $J_{\mathrm{E}}(f_1,f_1)$ and collision integrals demanding 
on interaction on each wall. Such a structure of $J(f_1,f_1)$ is caused by 
the structure of $\hat{T}$-operator:
%
%
\bea
J(f_1,f_1)=J_{\mathrm{E}}(f_1,f_1)+\sum_{i=1}^{n^*}[J_{lib}+J_{lic}+J_{lid}]+
\sum_{j=1}^{m^*}[J_{rjb}+J_{rjc}+J_{rjd}].\label{e4.2}
\eea
Going to the schema of construction of normal solutions to kinetic 
equations with the help of boundary conditions \cite{18}, let us introduce 
in the right-hand side of \refp{e4.1} an infinitely small source 
$-\bar\veps(f_1-f_1^{(0)})$, where $\bar\veps\to0$:
%
%
\bea
\lp\ddt+\vec{v}_1\dd{\vec{r}_1}\rp 
f_1(x_1;t)=J(f_1,f_1)-\bar\veps\lp f_1-f_1^{(0)}\rp.\label{e4.3}
\eea
For deviation $\delta f=f_1-f_1^{(0)}$ this equation reads:
%
%
\bea
&&\lp\ddt+\vec{v}_1\dd{\vec{r}_1}+\bar\veps\rp\delta f=
-\lp\ddt+\vec{v}_1\dd{\vec{r}_1}\rp f_1^{(0)}+{}\label{e4.4}\\
&&J\lp f_1^{(0)},f_1^{(0)}\rp+J\lp f_1^{(0)},\delta f\rp+
J\lp\delta f,f_1^{(0)}\rp+J\lp\delta f,\delta f\rp.\nonumber
\eea
First and foremost it should be noted that in the collision integral in the 
usual Boltzmann equation is local ($f_1$ is a function of the same 
Cartesian coordinate $\vec{r}_1$). In our case the $J(f_1,f_1)$ is a 
nonlocal collision integral. One time $f_1$ is calculated in $\vec{r}_1$, 
second time -- in $\vec{r}_1\pm\hs\sigma$, integration with respect to 
$\hs$ (surface of unit sphere) is performed. As a consequence of that in 
one or another way we will find solutions in some approximation and there 
is no sense ``to draw'' the whole nonlocal collision integral. It is much 
convenient to use its approximate expression. This approximation should be 
the same order as within the frame for the solution of total kinetic 
equation. In such a way, let us expand functions $f_1$ and 
$\g_2^{\mathrm{q}}$ in the vicinity of $\vec{r}_1$ into series on 
deviations $\pm\hs\sigma$. This results in:
%
%
\bea
\lefteqn{\ds J\equiv\sum_{k=0}^\infty J_k=J_0+J^*,}
\qquad\qquad\qquad\qquad\qquad&&
J^*=\sum_{k=1}^\infty J_k,\label{e4.5}\\
\lefteqn{\ds I\equiv\sum_{k=0}^\infty I_k=I_0+I^*,}
\qquad\qquad\qquad\qquad\qquad&&
I^*=\sum_{k=1}^\infty I_k,\label{e4.6}
\eea
where $k$ is the expansion order. All $J_k$ are here local functionals of 
$f_1$. We also use the following notations: 
$I(\delta f)=J\lp f_1^{(0)},\delta f\rp+J\lp\delta f,f_1^{(0)}\rp$ -- 
linearized nonlocal collision operator. We have similar expansion for this 
operator -- the relation \refp{e4.6}. Here 
$I_0(\delta f)=J_0\lp f_1^{(0)},\delta f\rp+J_0\lp\delta f,f_1^{(0)}\rp$ is 
linearized local collision operator which coincides with those one of the 
usual Boltzmann kinetic equation within a factor of 
$\g_2^{\mathrm{eq}}(\sigma_0^+)$ if one neglects the set of walls except 
hard sphere wall only. Then equation \refp{e4.4} transfers to
%
%
\bea
&&\ddt\delta f+\bar\veps\delta f-I_0(\delta f)={}\label{e4.7}\\
&&-\frac{\mathrm{D}}{\mathrm{D}t}f_1^{(0)}+
J\lp f_1^{(0)},f_1^{(0)}\rp+I^*(\delta f)
+J(\delta f,\delta f)-\vec{v}_1\dd{\vec{r}_1}\delta f.\nonumber
\eea
To solve this equation by means the boundary conditions method we need in 
its integral form. To this end let us introduce operator $S(t,t')$ with the 
following properties:
%
%
\bea
\ddt S(t,t')=I_0S(t,t'),\quad S(t,t)=1.\label{e4.8}
\eea
Using the limiting condition $\ds\lim_{t\to-\infty}\delta f(t)=0$ one 
obtains:
%
%
\bea
\delta f(t)&=&\int_{-\infty}^t\d t'\;\e^{\bar\veps(t'-t)}S(t,t')\times{}
\label{e4.9}\\
&&\ls\frac{\mathrm{D}}{\mathrm{D}t}f_1^{(0)}+
J\lp f_1^{(0)},f_1^{(0)}\rp+I^*(\delta f)+J(\delta f,\delta f)-
\vec{v}_1\dd{\vec{r}_1}\delta f\rs_{t'}.\nonumber
\eea
Equation \refp{e4.9} is completely ready for iteration procedure. This 
procedure can be organized as follows:
%
%
\bea
\delta f^{(k+1)}(t)&=&\int_{-\infty}^t\d t'\;\e^{\bar\veps(t'-t)}
S(t,t')\ls\frac{\mathrm{D}}{\mathrm{D}t}f_1^{(0)}+
J^{(k+1)}\lp f_1^{(0)},f_1^{(0)}\rp+{}\right.\label{e4.10}\\
&&\left.I^{*(k+1)}\lp\delta f^{(k)}\rp+
J^{(k+1)}\lp\delta f^{(k)},\delta f^{(k)}\rp-
\vec{v}_1\dd{\vec{r}_1}\delta f^{(k)}\rs_{t'},\nonumber
\eea
where
\bean
J^{(k+1)}&=&\sum_{k'=0}^{k+1}J_{k'},\\
I^{*(k+1)}&=&\sum_{k'=0}^{k+1}I_{k'}.\\
\eean
Each $(k+1)$th step uses conservation laws in $k$th approximation.

\section{Distribution function $\boldsymbol{f_1}$ in the first 
approximation}

The expression for distribution function $f_1$ in the first approximation 
is obtained if one puts into \refp{e3.5} $k=0$ and take into account the 
equality $\delta f^{(0)}=0$. Then we have:
%
%
\bea
\delta f^{(1)}(t)&=&\!\int_{-\infty}^t\d t'\;\e^{\veps(t'-t)}
S(t,t')\!\ls\frac{\mathrm{D}}{\mathrm{D}t}f_1^{(0)}+
J_0\lp f_1^{(0)},f_1^{(0)}\rp+J_1\lp f_1^{(0)},f_1^{(0)}\rp\rs_{t'}.
\nonumber\\\label{e5.1}
\eea
It can be shown that
%
%
\bea
J_0\lp f_1^{(0)},f_1^{(0)}\rp=0.\label{e5.2}
\eea
Making the expansion with an accuracy of first order on gradients of 
hydrodynamic parameters
%
%
\bea
\g_2^{\mathrm{q}}(\vec{r}_1,\vec{r}_2|n(t),\beta(t))&\simeq&
\g_2^{\mathrm{eq}}\lp r_{12};n\lp\frac{\vec{r}_1+\vec{r}_2}{2};t\rp,
\beta\lp\frac{\vec{r}_1+\vec{r}_2}{2};t\rp\rp\nonumber\\
&=&\frac12\vec{r}_{12}\dd{\vec{r}_1}\g_2^{\mathrm{eq}}+\ldots,\label{e5.3}\\
f_1(\vec{r}_1\pm\hs\sigma)&=&f_1(\vec{r}_1)\pm
\dd{\vec{r}_1}f_1\;\hs\sigma+\ldots\nonumber
\eea
and taking into consideration conservation laws \refp{e3.18} one obtains 
after very unwieldy calculations the following:
%
%
\bea
-\frac{\mathrm{D}}{\mathrm{D}}f_1^{(0)}+J_1\lp f_1^{(0)},f_1^{(0)}\rp=
K_\alpha\dd{r_{1\alpha}}\ln T+L_{\alpha\beta}\dd{r_{1\beta}}u_\alpha,
\label{e5.4}
\eea
%
%
%
%
%
\bea
\lefteqn{K_\alpha}\phantom{L_{\alpha\beta}}
&=&-f_1^{(0)}\ls1+\frac35\frac{p^{\mathrm{i}}}{n\kB T}\rs
\ls\frac{mc_1^2}{2\kB T}-\frac52\rs c_{1\alpha}+
\sum_{i=1}^{n^*}K_{\alpha li}+\sum_{j=1}^{m^*}K_{\alpha rj},\label{e5.5}\\
\lefteqn{L_{\alpha\beta}}\phantom{L_{\alpha\beta}}
&=&-f_1^{(0)}
\ls1+\frac25\frac{p^{\mathrm{i}}}{n\kB T}\rs \frac{m}{\kB T}
\ls c_{1\alpha}c_{1\beta}-\frac13c_1^2\delta_{\alpha\beta}\rs +
\sum_{i=1}^{n^*}L_{\alpha\beta li}+\sum_{j=1}^{m^*}L_{\alpha\beta rj},
\nonumber\\\label{e5.6}
\eea
where
%
%
\bea
K_{\alpha{li\atop rj}}&=&\mp\frac8{\pi^2}n^2\sigma_{{li\atop rj}}
\g_2^{\mathrm{eq}}\lp\sigma_{li\atop rj}^{\pm}\rp
\frac{m}{2\kB T}\times{}\label{e5.7}\\
&&\ls\frac1{30}\e^{-s^2}\int\d\vec{x}_2\;\exp\lc-h^2-v^2\rc h_\beta
\lp v_\alpha v_\beta-\frac13 v^2\delta_{\alpha\beta}\rp\times{}\right.
\nonumber\\
&&\phantom{\bigg[}\lc4+\frac{s^3}{v^5}\lp5v^2+6s^2\rp+\frac{1}{v^5}
\lp v^2+s^2\rp^{3/2}\lp4v^2-6s^2\rp\rc-{}\nonumber\\
&&\phantom{\bigg[}
\frac1{30}\int_{v>s}\d\vec{x}_2\;\exp\lc-h^2-v^2\rc h_\beta
\lp v_\alpha v_\beta-\frac13 v^2\delta_{\alpha\beta}\rp\times{}\nonumber\\
&&\phantom{\bigg[}\lc4+\frac{s^3}{v^5}\lp5v^2-9s^2\rp+\frac{1}{v^5}
\lp v^2-s^2\rp^{3/2}\lp4v^2+6s^2\rp\rc+{}\nonumber\\
&&\phantom{\bigg[}
\frac19\e^{-s^2}\int\d\vec{x}_2\;\exp\lc-h^2-v^2\rc h_\alpha
\lp v^2-\frac{s^3}{v}+\frac1v\lp v^2+s^2\rp^{3/2}\rp-{}
\nonumber\\
&&\phantom{\bigg[}\frac19
\int_{v>s}\left.\d\vec{x}_2\;\exp\lc-h^2-v^2\rc h_\alpha
\lp v^2-\frac{s^3}{v}+\frac1v\lp v^2+s^2\rp^{3/2}\rp\rs,\nonumber
\eea
%
%
%
%
%
\bea
L_{\alpha\beta{li\atop rj}}&=&\mp\frac8{\pi^2}n^2\sqrt2
\sigma_{{li\atop rj}}
\g_2^{\mathrm{eq}}\lp\sigma_{li\atop rj}^{\pm}\rp\lp
\frac{m}{2\kB T}\rp^{3/2}\times{}
\label{e5.8}\\
&&\ls\frac1{30}\e^{-s^2}\int\d\vec{x}_2\;\exp\lc-h^2-v^2\rc
\lp v_\alpha v_\beta-\frac13 v^2\delta_{\alpha\beta}\rp\times{}\right.
\nonumber\\
&&\phantom{\bigg[}\lc4+\frac{s^3}{v^5}\lp5v^2+6s^2\rp+\frac{1}{v^5}
\lp v^2+s^2\rp^{3/2}\lp4v^2-6s^2\rp\rc-{}\nonumber\\
&&\phantom{\bigg[}\frac1{30}\int_{v>s}\d\vec{x}_2\;\exp\lc-h^2-v^2\rc
\lp v_\alpha v_\beta-\frac13 v^2\delta_{\alpha\beta}\rp\times{}\nonumber\\
&&\phantom{\bigg[}\lc4+\frac{s^3}{v^5}\lp5v^2-9s^2\rp+\frac{1}{v^5}
\lp v^2-s^2\rp^{3/2}\lp4v^2+6s^2\rp\rc+{}\nonumber\\
&&\phantom{\bigg[}\frac19\e^{-s^2}\int\d\vec{x}_2\;\exp\lc-h^2-v^2\rc
\lp v^2-\frac{s^3}{v}+\frac1v\lp v^2+s^2\rp^{3/2}\rp\delta_{\alpha\beta}-{}
\nonumber\\
&&\phantom{\bigg[}\frac19
\int_{v>s}\left.\d\vec{x}_2\;\exp\lc-h^2-v^2\rc
\lp v^2-\frac{s^3}{v}+\frac1v\lp v^2+s^2\rp^{3/2}\rp\delta_{\alpha\beta}
\rs.\nonumber
\eea
Here
%
%
\bea
s=\lc
\ba{l}
\lp\beta\vt\veps_{li}\rp^{1/2},\\
\lp\beta\vt\veps_{rj}\rp^{1/2},\\
\ea\right.\quad\vec{x}_2=\lp\frac{m}{4\kB T}\rp^{1/2}\vec{c}_2,\label{e5.9}
\eea
%
%
%
%
%
\bea
\ba{lclcl}
\ds\vec{v}=\lp\frac{m}{4\kB T}\rp^{1/2}\gb,&\quad&
\gb&=&\vec{c}_2-\vec{c}_1,\\
\ds\vec{h}=2\lp\frac{m}{4\kB T}\rp^{1/2}\vec{G}_0,&\quad&
\vec{G}_0&=&\ds\frac12(\vec{c}_1+\vec{c}_2).\\
\ea\label{e5.10}
\eea
Then
%
%
\bea
f_1\equiv f_1^{(1)}&=&f_1^{(0)}+\delta f^{(1)},\label{e5.11}\\
\delta f^{(1)}(t)&=&\int_{-\infty}^t\d t'\;\e^{\veps(t'-t)}S(t,t')
\lc K_\alpha\dd{r_{1\alpha}}\ln T+
L_{\alpha\beta}\dd{r_{1\beta}}u_\alpha\rc_{t'}.\nonumber
\eea
It can be shown that 
$\int\d\vec{v}_1\;\delta f^{(1)}(\vec{r}_1,\vec{v}_1;t)\vec{\Psi}=0$, i.e. 
the hydrodynamic parameters $n\rt$, $\vec{u}\rt$, $\beta$, 
$\omega_{\mathrm{i}}$ are completely defined by the local one-particle 
distribution function $f_1^{(0)}$ \refp{e3.14}.

\section{Conservation laws in the first approximation. Stationary process}

First, let us calculate kinetic parts of the stress tensor and heat flux 
vector. Substituting one-particle distribution function $f_1$ \refp{e5.11} 
into \refp{e3.5} one obtains:
%
%
\bea
P_{\alpha\beta}^{\mathrm{k}(1)}&=&p^{\mathrm{k}}\delta_{\alpha\beta}+
\int\d t'\;\e^{\veps(t'-t)}M^{\mathrm{k}}(t,t')
\ls S_{\alpha\beta}\rs_{t'},\label{e6.1}\\
S_{\alpha\beta}&=&\dd{r_{1\alpha}}u_{\alpha}+\dd{r_{1\beta}}u_{\beta}-\frac23
\dd{r_{1\gamma}}u_{\gamma}\delta_{\alpha\beta},\nonumber\\
q_{\alpha}^{\mathrm{k}(1)}&=&\int\d t'\;\e^{\veps(t'-t)}L^{\mathrm{k}}(t,t')
\ls\dd{r_{1\alpha}}\ln T\rs_{t'},\label{e6.2}
\eea
where cores of kinetic part of transport laws read:
%
%
\bea
M^{\mathrm{k}}(t,t')&=&\frac{1}{10}\int\d\vec{v}_1\;
mc_{1\alpha}c_{1\beta}S(t,t')\lc L_{\alpha\beta}\rc_{t'},\label{e6.3}\\
L^{\mathrm{k}}(t,t')&=&\;\frac13\,\int\d\vec{v}_1\;
c_{1\alpha}\frac{mc_1^2}{2}S(t,t')\lc K_\alpha\rc_{t'}.\label{e6.4}
\eea
For calculation of potential (interaction) parts of 
$P_{\alpha\beta}^{\mathrm{i}(1)}$ and $q_\alpha^{\mathrm{i}(1)}$, the 
expression $\int_0^1\d\lambda f_2(\vec{r}_1+\lambda\hs\sigma,\vec{v}_1,
\vec{r}_1+\lambda\hs\sigma-\hs\sigma,\vec{v}_2;t)=z_1+z_2$ should be 
expanded into series on, first of all, inhomogeneity of distribution 
function, then on deviation of $f_1^{(0)}$ on $\delta f^{(1)}$. In both 
cases one should keep only the first order on gradients. Calculations 
give:
%
%
\bea
z_1&=&\frac12\sigma\g_2^{\mathrm{eq}}(\sigma)f_1^{(0)}(x_1;t)
f_1^{(0)}(\vec{r}_1,\vec{v}_2;t)\hs\dd{\vec{r}_1}
\ln\frac{f_1^{(0)}(\vec{r}_1,\vec{v}_1;t)}
{f_1^{(0)}(\vec{r}_1,\vec{v}_2;t)},\label{e6.5}\\
z_2&=&\g_2^{\mathrm{eq}}(\sigma)\lc f_1^{(0)}(x_1;t)
\delta f^{(1)}(\vec{r}_1,\vec{v}_2;t)+
\delta f^{(1)}(x_1;t)f_1^{(0)}(\vec{r}_1,\vec{v}_2;t)\rc.\label{e6.6}
\eea
$z_1$ is the expansion of $f_1^{(0)}$ on inhomogeneity (the inhomogeneity 
of $\delta f^{(1)}$ is considered as a negligibly small quantity), $z_2$ is 
the expansion on deviation $\delta f^{(1)}$. Then, general expressions 
\refp{e3.6}--\refp{e3.13} taking into account \refp{e6.5} and \refp{e6.6} 
transfers to
%
%
\bea
P_{\alpha\beta}^{\mathrm{i}1}&=&\int\d t'\;\e^{\veps(t'-t)}
M^{\mathrm{i}}(t,t')\ls S_{\alpha\beta}\rs_{t'}-{}\label{e6.7}\\
&&\frac49n^2\sqrt{\pi m\kB T}\;H_2
\lc\frac65S_{\alpha\beta}+\dd{r_{1\gamma}}u_\gamma
\delta_{\alpha\beta}\rc,\nonumber\\
q_\alpha^{\mathrm{i}1}&=&\int\d t'\;\e^{\veps(t'-t)}
L^{\mathrm{i}}(t,t')\ls\dd{r_{1\alpha}}\ln T\rs_{t'}-
\frac23n^2\kB\sqrt{\frac{\pi\kB T}{m}}H_2
\frac{\p T}{\p r_{1\alpha}},\label{e6.8}
\eea
where $M^{\mathrm{i}}$, $L^{\mathrm{i}}$ are cores of potential parts of 
transfer laws. Their structure is very complicated. To save space, these 
expressions are not presented here in their explicit form.
%
%
\bea
H_2=\sigma_0^4\g_2^{\mathrm{eq}}(\sigma_0^+)&+&
\sum_{i=1}^{n^*}\sigma_{li}^4\g_2(\sigma_{li}^+)\e^{-\beta\vts\veps_{li}}
\Xi(\beta\vt\veps_{li})\label{e6.9}\\
&+&\sum_{j=1}^{m^*}\sigma_{rj}^4\g_2^{\mathrm{eq}}(\sigma_{rj}^+)
\e^{-\beta\vts\veps_{rj}}\Xi(\beta\vt\veps_{rj}),\nonumber
\eea
%
%
%
%
%
\bea
\ba{lcl}
\Xi(s)&=&\ds\e^s-\frac12s-K_1^*(s),\\ [1ex]
K_1^*(s)&=&\ds2\int\nolimits_0^\infty\d x\;x^2\sqrt{x^2+s}.\\
\ea\label{e6.10}
\eea

In such a way, transport laws in the first approximation are not 
completely integral as this takes place in the case of solution of the 
usual Boltzmann kinetic equation \cite{18}. There are also local-time terms 
(second terms in \refp{e6.7} and \refp{e6.8}). They are caused by the 
inhomogeneity of $f_1^{(0)}$ and, consequently, are not sensitive to the 
``memory'' effects.

In stationary case operator $I_0$ does not depend on time explicitly:
\bean
S(t,t')=\e^{I_0(t-t')}.
\eean
Then
%
%
\bea
\delta f^{(1)}=\int\d\tau\;\e^{\veps\tau}\e^{-I_0\tau}
\lc K_\alpha\dd{r_{1\alpha}}\ln T+
L_{\alpha\beta}\dd{r_{1\beta}}u_\alpha\rc,\quad\tau=t'-t.\label{e6.11}
\eea
Let us define the following quantities:
%
%
\bea
\ba{lclclcl}
a_\alpha(\tau)&=&\ds\e^{-I_0\tau}\lc K_\alpha\dd{r_{1\alpha}}\ln T\rc,
&\quad&a_\alpha(0)&=&\ds K_\alpha\dd{r_{1\alpha}}\ln T,\\ [1.5ex]
b_{\alpha\beta}(\tau)&=&\ds\e^{-I_0\tau}
\lc L_{\alpha\beta}\dd{r_{1\beta}}u_\alpha\rc,
&\quad&b_{\alpha\beta}(0)&=&\ds L_{\alpha\beta}\dd{r_{1\beta}}u_\alpha.\\
\ea\label{e6.12}
\eea
It can be shown that the operator $I_0$ has the same mathematical 
properties that the corresponding operator of the usual Boltzmann kinetic 
equation:
%
%
\bea
I_0\lp f_1^{(0)}\vec{\Psi}\rp=0,\quad I_0(\xi)=\lambda\xi,\quad\lambda<0.
\label{e6.13}
\eea
Then
%
%
\bea
\ba{lcll}
\|a_\alpha(\tau)\|&\leqslant&\|a_\alpha(0)\|&\exp\{\lambda_{\max}\tau\},\\
\|b_{\alpha\beta}(\tau)\|&\leqslant&
\|b_{\alpha\beta}(0)\|&\exp\{\lambda_{\max}\tau\},\\
\ea\quad\tau<0,\quad\lambda_{\max}\equiv\max\{\lambda\}\label{e6.14}
\eea
and
\bean
\|\vphi\|=\int\d\vec{v}_1\;\ls f^{(0)}\rs^{-1}\vphi^2.
\eean
Using last transformations, equation \refp{e6.11} transfers to:
%
%
\bea
\delta f^{(1)}=\lim_{\veps\to+0}\int_{-\infty}^0\d\tau\;\e^{\veps\tau}
\Big(a_\alpha(\tau)+b_{\alpha\beta}(\tau)\Big)=\int_{-\infty}^0\d\tau\;
\Big(a_\alpha(\tau)+b_{\alpha\beta}(\tau)\Big),\label{e6.15}
\eea
or, introducing $\delta f^{(1)}=\phi^{(1)}f_1^{(0)}$, it can be rewritten 
in the following final form:
%
%
\bea
\phi^{(1)}=A_\alpha\dd{r_{1\alpha}}\ln T+B_{\alpha\beta}
\dd{r_{1\beta}}u_\alpha.\label{e6.16}
\eea
$A_\alpha$ and $B_{\alpha\beta}$ satisfy integral equations like this:
%
%
\bea
\ba{lcl}
I_0(A_\alpha)&=&K_\alpha,\\
I_0(B_{\alpha\beta})&=&L_{\alpha\beta}.\\
\ea\label{e6.17}
\eea
Operator $I_0$ has the following structure:
%
%
\bea
I_0=I_a+\sum_{i=1}^{n^*}\lc I_{ib}+I_{ic}+I_{id}\rc+
\sum_{j=1}^{m^*}\lc I_{jb}+I_{jc}+I_{jd}\rc,\label{e6.18}
\eea
where
%
%
\bea
I_a(\phi_1)&=&\sigma_0^2\g_2^{\mathrm{eq}}(\sigma_0^+)
\int\d\vec{v}_2\d\hs\;\hs\gb\;
\thsgb\times{}\label{e6.19}\\
&&f_1^{(0)}(\vec{v}_1)f_1^{(0)}(\vec{v}_2)
\lc\phi_1+\phi_2-\phi_1'-\phi_2'\rc,\nonumber\\
I_{ib}(\phi_1)&=&\sigma_{li}^2\g_2^{\mathrm{eq}}(\sigma_{li}^-)
\int\d\vec{v}_2\d\hs\;\hs\gb\;\tmhsgb\times{}\label{e6.20}\\
&&f_1^{(0)}(\vec{v}_1)f_1^{(0)}(\vec{v}_2)
\lc\phi_1+\phi_2-\phi_{1i}''-\phi_{2i}''\rc,\nonumber\\
I_{ic}(\phi_1)&=&\sigma_{li}^2\g_2^{\mathrm{eq}}(\sigma_{li}^+)
\int\d\vec{v}_2\d\hs\;\hs\gb\;\theta\lp\hs\gb-
\sqrt{\frac{4\vt\veps_{li}}{m}}\,\rp\times{}\label{e6.21}\\
&&f_1^{(0)}(\vec{v}_1)f_1^{(0)}(\vec{v}_2)
\lc\phi_1+\phi_2-\phi_{1i}'''-\phi_{2i}'''\rc,\nonumber\\
I_{id}(\phi_1)&=&\sigma_{li}^2\g_2^{\mathrm{eq}}(\sigma_{li}^+)
\int\d\vec{v}_2\d\hs\;\hs\gb\;\thsgb
\theta\lp\sqrt{\frac{4\vt\veps_{li}}{m}}-\hs\gb\rp\times{}\label{e6.22}\\
&&f_1^{(0)}(\vec{v}_1)f_1^{(0)}(\vec{v}_2)
\lc\phi_1+\phi_2-\phi_{1i}''''-\phi_{2i}''''\rc,\nonumber\\
I_{jb}(\phi_1)&=&\sigma_{rj}^2\g_2^{\mathrm{eq}}(\sigma_{rj}^+)
\int\d\vec{v}_2\d\hs\;\hs\gb\;\thsgb\times{}\label{e6.23}\\
&&f_1^{(0)}(\vec{v}_1)f_1^{(0)}(\vec{v}_2)
\lc\phi_1+\phi_2-\phi_{1j}''-\phi_{2j}''\rc,\nonumber\\
I_{jc}(\phi_1)&=&\sigma_{rj}^2\g_2^{\mathrm{eq}}(\sigma_{rj}^-)
\int\d\vec{v}_2\d\hs\;\hs\gb\;\theta\lp-\hs\gb-
\sqrt{\frac{4\vt\veps_{rj}}{m}}\,\rp\times{}\label{e6.24}\\
&&f_1^{(0)}(\vec{v}_1)f_1^{(0)}(\vec{v}_2)
\lc\phi_1+\phi_2-\phi_{1j}'''-\phi_{2j}'''\rc,\nonumber\\
I_{jd}(\phi_1)&=&\sigma_{rj}^2\g_2^{\mathrm{eq}}(\sigma_{rj}^+)
\int\d\vec{v}_2\d\hs\;\hs\gb\;\tmhsgb
\theta\lp\sqrt{\frac{4\vt\veps_{rj}}{m}}+\hs\gb\rp\times{}\label{e6.25}\\
&&f_1^{(0)}(\vec{v}_1)f_1^{(0)}(\vec{v}_2)
\lc\phi_1+\phi_2-\phi_{1j}''''-\phi_{2j}''''\rc,\nonumber
\eea
%
%
%
%
%
\bea
\ba{lclclcl}
\phi_1&\equiv&\phi(\vec{r}_1,\vec{v}_1;t),&\quad&
\phi_{1{i\atop j}}&\equiv&\phi\lp\vec{r}_1,\vec{v}_{1{l\atop r}}^*;t\rp,\\ 
[1ex]
\phi_2&\equiv&\phi(\vec{r}_1,\vec{v}_2;t),&\quad&
\phi_{2{i\atop j}}&\equiv&\phi\lp\vec{r}_1,\vec{v}_{2{l\atop r}}^*;t\rp,\\
\ea\quad*\equiv(','',''','''').\label{e6.26}
\eea
In the case of SET (RET) theory, the linearized local integral operator 
$I_0=I_a$, whereas in the case of the Boltzmann kinetic equation there is 
usual Boltzmann's linearized operator and $I_a$ tends to that one in the 
limit of low density: $n\to0$, $\g_2^{\mathrm{eq}}(\sigma_0^+)\to1$, 
$I_a\to I_{\mathrm{B}}$.

In such a way, to find one-particle distribution function in the first 
approximation in stationary case one should analyse integral equations 
\refp{e6.17} and solve them.

\section{Solutions to the integral equations}

To find quantities $A_\alpha$ and $B_{\alpha\beta}$ we have set of integral 
equations \refp{e6.17}. Let us define dimensionless self velocity: 
$\vec{w}=\lp\frac{m}{2\kB T}\rp^{1/2}\vec{c}$. Using the property of 
isotropy (in the velocity space) of the operator $I_0$ and structures of 
$K_\alpha$ \refp{e5.7} and $L_{\alpha\beta}$ \refp{e5.8} solutions to 
\refp{e6.17} can be presented as follows:
%
%
\bea
A_\alpha&=&w_{1\alpha}A(w_1),\label{e7.1}\\
B_{\alpha\beta}&=&B_{1\alpha\beta}+B_{2\alpha\beta}(w_1)
\delta_{\alpha\beta},\label{e7.2}\\
B_{1\alpha\beta}&=&\ds\lp w_{1\alpha}w_{1\beta}-\frac13w_1^2
\delta_{\alpha\beta}\rp B_1(w_1).\nonumber
\eea
The structure of $B_{\alpha\beta}$ is caused by the structure of 
$L_{\alpha\beta}$: $L_{\alpha\beta}=L_{1\alpha\beta}+L_{2\alpha\beta}$, 
where in $L_{1\alpha\beta}$ are all terms with 
$\lp w_{1\alpha}w_{1\beta}-\frac13 w_1^2\delta_{\alpha\beta}\rp$, in 
$L_{2\alpha\beta}$ are all terms with $\delta_{\alpha\beta}$. Then
%
%
\bea
&&I_0(w_{1\alpha}A(w_1))=K_\alpha,\nonumber\\
&&I_0\lp\lp w_{1\alpha}w_{1\beta}-\frac{1}{3}w_1^2\delta_{\alpha\beta}\rp
B_1(w_1)\rp=L_{1\alpha\beta},\label{e7.3}\\
&&I_0(B_2(w_1))=L_2.\nonumber
\eea
Following the standard Chapman-Enskog method \cite{19}, let us represent 
$A(w_1)$, $B_1(w_1)$ and $B_2(w_1)$ via the Sonine-Laguerre polynomials
%
%
\bea
S_n^m(z)=\sum_{j=0}^m(-z)^j\frac{\Gamma(n+m+1)}{j!(m-j)!\Gamma(n+j+1)},
\label{e7.4}
\eea
i.e in the form
%
%
\bea
\ba{lcl}
A(w_1)&=&\ds\sum_{m=0}^\infty a^{(m)}S_{3/2}^{(m)}(w_1^2),\\
B_1(w_1)&=&\ds\sum_{m=0}^\infty b_1^{(m)}S_{5/2}^{(m)}(w_1^2),\\
B_2(w_1)&=&\ds\sum_{m=0}^\infty b_2^{(m)}S_{1/2}^{(m)}(w_1^2).\\
\ea\label{e7.5}
\eea
Fredgolm condition gives limitations on expansion coefficients $a^{(m)}$ 
and $b^{(m)}$: the correction in the first approximation does not 
contribute to hydrodynamic parameters 
$\int\d\vec{w}_1\;f_1^{(0)}\phi^{(1)}=0$. By this means:
\bea
a^{(0)}&=&0,\nonumber\\
b^{(0)}&=&0,\label{e7.6}\\
b^{(1)}&=&0.\nonumber
\eea
It is known that the Sonine-Laguerre polynomials converges quickly. 
Therefore only the first nonzero term is considered in expansion. This is 
as a rule and we will follow the procedure. Such an approximation gives the 
error for transport coefficients (practically for all types of interaction 
potentials) not exceeding 2\%. Thus, we have:
%
%
\bea
\ba{lclcl}
A(w_1)&\simeq&a^{(1)}S_{3/2}^{(1)}(w_1^2)&=&
\ds a^{(1)}\lp\frac52-w_1^2\rp,\\
B_1(w_1)&\simeq&b_1^{(0)}S_{5/2}^{(0)}(w_1^2)&=&b_1^{(0)},\\
B_2(w_1)&\simeq&b_2^{(0)}S_{1/2}^{(2)}(w_1^2)&=&\ds b_2^{(2)}
\lp\frac{15}{8}-\frac{5}{2}w_1^2+\frac12w_1^4\rp,\\
\ea\label{ee7.7}
\eea
and
%
%
\bea
&&I_0\lp w_{1\alpha}a^{(1)}\lp\frac52-w_1^2\rp\rp=K_\alpha,\nonumber\\
&&I_0\lp\lp w_{1\alpha}w_{1\beta}-\frac13w_1^2\delta_{\alpha\beta}\rp
b_1^{(0)}\rp=L_{1\alpha\beta},\label{e7.8}\\
&&I_0\lp b_2^{(2)}\lp\frac{15}{8}-\frac{5}{2}w_1^2+\frac12w_1^4\rp\rp=
L_2.\nonumber
\eea
Multiplying these equations on $w_{1\alpha}S_{3/2}^{(1)}(w_1^2)$, 
$\lp w_{1\alpha}w_{1\beta}-\frac13 w_1^2\delta_{\alpha\beta}\rp$, 
$S_{1/2}^{(2)}(w_1^2)$, correspondingly, and integrating with respect to 
$\vec{w}_1$, one finds:
%
%
\bea
a^{(1)}&=&\frac{\int\d\vec{w}_1\;K_\alpha w_{1\alpha}S_{3/2}^{(1)}(w_1^2)}
{\int\d\vec{w}_1\;I_0\lp w_{1\alpha}S_{3/2}^{(1)}(w_1^2)\rp
w_{1\alpha}S_{3/2}^{(1)}(w_1^2)},\label{e7.9}\\
b_1^{(0)}&=&\frac{\int\d\vec{w}_1\;L_{1\alpha\beta}
\lp w_{1\alpha}w_{1\beta}-\frac13 w_1^2\delta_{\alpha\beta}\rp}
{\int\d\vec{w}_1\;I_0\lp w_{1\alpha}w_{1\beta}-
\frac13 w_1^2\delta_{\alpha\beta}\rp
\lp w_{1\alpha}w_{1\beta}-\frac13 w_1^2\delta_{\alpha\beta}\rp},
\label{e7.10}\\
b_2^{(0)}&=&\frac{\int\d\vec{w}_1\;L_2S_{1/2}^{(2)}(w_1^2)}
{\int\d\vec{w}_1\;I_0\lp S_{1/2}^{(2)}(w_1^2)\rp
S_{1/2}^{(2)}(w_1^2)}.\label{e7.11}
\eea
First-hand calculations for $a^{(1)}$ and $b^{(0)}$ give the following:
%
%
\bea
a^{(1)}&=&\phantom{-}\frac{\frac{15}{4}n\lp1+\frac25\pi n\Lambda\rp-
\frac32\sqrt{\pi}n^2D_1}{8\sqrt{2\pi}n^2\lc\lambda^*+\frac{11}{32}D_2\rc},
\label{e7.12}\\
b_1^{(0)}&=&-\frac{5n\lp1+\frac{4}{15}\pi n\Lambda\rp-
\frac43\sqrt{\pi}n^2D_1}{8\sqrt{2\pi}n^2\lc\lambda^*+\frac{1}{12}D_2\rc}
\lp\frac{m}{2\kB T}\rp^{1/2},\label{e7.13}
\eea
where
\bean
\ba{lll}
D_1&=&-\ds\sum_{i=1}^{n^*}\sigma_{li}^3
\g_2^{\mathrm{eq}}(\sigma_{li}^+)\vt s_i
\e^{-\vts s_i}H_s(\vt s_i)\\
&&\ds+\sum_{j=1}^{m^*}\sigma_{rj}^3
\g_2^{\mathrm{eq}}(\sigma_{rj}^-)\vt s_j
\e^{-\vts s_j}H_s(\vt s_j),\\
D_2&=&\phantom{-}
\ds\sum_{i=1}^{n^*}\sigma_{li}^3\g_2^{\mathrm{eq}}(\sigma_{li}^+)\vt s_i^2
\e^{-\vts s_i}+
\sum_{j=1}^{m^*}\sigma_{rj}^3\g_2^{\mathrm{eq}}(\sigma_{rj}^-)\vt s_j^2
\e^{-\vts s_j},\\
\ea
\eean
\bean
\ba{lclclcl}
\vt s_i&=&\beta\vt\veps_{li},&\quad&i&=&1,\ldots,n^*,\\
\vt s_j&=&\beta\vt\veps_{rj},&\quad&j&=&1,\ldots,m^*,\\
\ea
\eean
%
%
%
%
%
\bea
&&\text{\small$\lambda^*=\frac12\!\ls\sigma_0\g_2^{\mathrm{eq}}(\sigma_0^+)+
\sum_{i=1}^{n^*}\sigma_{li}^3\g_2^{\mathrm{eq}}(\sigma_{li}^+)\e^{\vts s_i}
\Xi(\vt s_i)+
\sum_{j=1}^{m^*}\sigma_{rj}^3\g_2^{\mathrm{eq}}(\sigma_{rj}^-)\e^{\vts s_j}
\Xi(\vt s_j)\rs$}\!,\nonumber\\
&&H_s(z)=\sqrt{\pi}/2+\e^s\Gamma(3/2,z),\label{e7.14}
\eea
$\Gamma(r,s)=\int_s^\infty\d x\;x^{r-1}\e^{-x}$ is incomplete 
$\Gamma$-function. As far as $b_2^{(2)}$ does not contribute into transport 
coefficients, we did not calculate it.

In this manner, one-particle distribution function in stationary case reads:
%
%
\bea
f_1^{(1)}&=&f_1^{(0)}\lp1+\phi^{(1)}\rp,\label{e7.15}\\
\phi^{(1)}&\simeq&a^{(1)}\lp\frac52-w_1^2\rp w_{1\alpha}\dd{r_{1\alpha}}
\ln T+b_1^{(0)}\lp w_{1\alpha}w_{1\beta}-
\frac13w_1^2\delta_{\alpha\beta}\rp\dd{r_{1\beta}}u_\alpha\nonumber\\
&&\phantom{a^{(1)}\lp\frac52-w_1^2\rp w_{1\alpha}\dd{r_{1\alpha}}\ln T}+
b_2^{(2)}\lp\frac{15}{8}-\frac52w_1^2+\frac12w_1^4\rp
\dd{r_{1\alpha}}u_\alpha,\nonumber
\eea
where $a^{(1)}$ and $b^{(2)}$ are defined by \refp{e7.12} and \refp{e7.13}, 
correspondingly.

\section{Calculation of transport coefficients. Some limiting cases}

General expressions for the stress tensor and heat flux vector for 
nonstationary process were obtained in section 6. But explicit calculation 
was performed for only one part which is inhomogeneous on $f_1^{(0)}$ and 
local-time next to other integral terms. In stationary case integral terms 
transfers to local-time. The explicit calculation for these terms becomes 
possible. Substituting one-particle distribution function $f_1$ 
\refp{e7.15} into general expressions \refp{e3.5}--\refp{e3.13} and taking 
into account new structure for \refp{e6.6}:
%
%
\bea
z_2=\g_2^{\mathrm{eq}}(\sigma)f_1^{(0)}(x_1;t)
f_1^{(0)}(\vec{r}_1,\vec{v}_2;t)
\lc\phi^{(1)}(x_1;t)+\phi^{(2)}(\vec{r}_1,\vec{v}_2;t)\rc,\label{e8.1}
\eea
one obtains:
%
%
\bea
P_{\alpha\beta}&=&p\delta_{\alpha\beta}-
\kappa\dd{r_{1\gamma}}u_\gamma\delta_{\alpha\beta}-2\eta S_{\alpha\beta},
\label{e8.2}\\
q_\alpha&=&-\lambda\dd{r_{1\alpha}}T,\label{e8.3}
\eea
where $S_{\alpha\beta}$ is the velocities shift tensor. Explicit 
expressions for transport coefficients bulk $\kappa$ and shear $\eta$ 
viscosities and thermal conductivity $\lambda$ read:
%
%
\bea
\kappa&=&\frac49n^2\lp\pi m\kB T\rp^{1/2}H_2,\label{e8.4}\\
\eta&=&\frac35\kappa+\frac12n\kB T\lc1+\frac{8}{15}\sqrt{\pi}nH_1\rc 
b^{(0)},\label{e8.5}\\
\lambda&=&\frac{3\kB}{2m}\kappa+\frac54n\kB\lp\frac{2\kB T}{m}\rp^{1/2}
\lc1+\frac45\sqrt{\pi}nH_1\rc a^{(1)},\label{e8.6}
\eea
where
%
%
\bea
&&b^{(0)}=-b_1^{(0)},\quad H_1=\frac{\sqrt{\pi}}{2}\Lambda-{}\label{e8.7}\\
&&\sum_{i=1}^{n^*}\sigma_{li}^3
\g_2^{\mathrm{eq}}(\sigma_{li}^+)\vt s_i\e^{-\vts s_i}
\lc\frac{\sqrt{\pi}}{4}-\frac{\vt s_i^{3/2}}{3}+
\frac13\e^{\vts s_i}\Gamma\lp\frac52,\vt s_i\rp\rc+{}\nonumber\\
&&\sum_{j=1}^{m^*}\sigma_{rj}^3
\g_2^{\mathrm{eq}}(\sigma_{rj}^-)\vt s_j\e^{-\vts s_j}
\lc\frac{\sqrt{\pi}}{4}-\frac{\vt s_j^{3/2}}{3}+
\frac13\e^{\vts s_j}\Gamma\lp\frac52,\vt s_j\rp\rc.\nonumber
\eea
Thus, the problem of transport coefficients for specific interaction 
potential \refp{e2.2} in our approach is solved. In concluding of this 
section let us consider some limiting cases.

\subsection{Hard spheres potential}

%
%
\bea
\ba{lclclclclcl}
\vt\veps_{li}&=&0,&\quad&
\vt s_i&=&0,&\quad&i&=&1,\ldots,n^*,\\
\vt\veps_{rj}&=&0,&\quad&
\vt s_j&=&0,&\quad&j&=&1,\ldots,m^*.\\
\ea\label{e8.8}
\eea
In this case model MSPI \refp{e2.2} transfers to that one for hard spheres, 
whereas kinetic equation \refp{e2.13} transfers to that one of the SET 
(RET) theory. It is naturally to expect that results 
\refp{e8.4}--\refp{e8.6} should transfer to the well known results of the 
SET theory. This assertion really takes place. With the condition 
\refp{e8.8} we have:
\bean
\ba{lclclcl}
\Lambda&\to&\sigma_0^3\g_2^{\mathrm{eq}}(\sigma_0^+),&\quad&D_1&\to&0,\\
H_1&\to&\ds\frac{\sqrt{\pi}}{2}\Lambda,&\quad&D_2&\to&0,\\
H_2&\to&\sigma_0^4\g_2^{\mathrm{eq}}(\sigma_0^+),&\quad&\Xi(0)&=&0,\\
H_2&\to&\sqrt{\pi},&\quad&K_1^*(0)&=&1,\\
\lambda^*&\to&\ds\frac12\sigma_0^2\g_2^{\mathrm{eq}}(\sigma_0^+),&\quad&
\Gamma(3/2)&=&\ds\frac{\sqrt{\pi}}{2}.\\
\ea
\eean
Then
%
%
\bea
a^{(1)}&\to&\frac{\frac{15}{4}n\lp1+
\frac25\pi n\sigma_0^3\g_2^{\mathrm{eq}}(\sigma_0^+)\rp}
{4\sqrt{2\pi}n^2\sigma_0^2\g_2^{\mathrm{eq}}(\sigma_0^+)},\label{e8.9}\\
b^{(0)}&\to&\frac{5n\lp1+
\frac{4}{15}\pi n\sigma_0^3\g_2^{\mathrm{eq}}(\sigma_0^+)\rp}
{4\sqrt{2\pi}n^2\sigma_0^2\g_2^{\mathrm{eq}}(\sigma_0^+)}
\lp\frac{m}{2\kB T}\rp^{1/2},
\label{e8.10}
\eea
%
%
%
%
%
\bea
p&\to&n\kB T\lp1+\frac23\pi n\sigma_0^3
\g_2^{\mathrm{eq}}(\sigma_0^+)\rp,\label{e8.11}\\
\kappa&\to&\frac49n^2\sqrt{\pi m\kB T}\sigma_0^4
\g_2^{\mathrm{eq}}(\sigma_0^+),\label{e8.12}\\
\eta&\to&\frac35\kappa+\frac{5}{16}\lp\frac{m\kB T}{\pi}\rp^{1/2}
\frac{1}{\sigma_0\g_2^{\mathrm{eq}}(\sigma_0^+)}\lp1+
\frac{4}{15}\pi n\sigma_0^3\g_2^{\mathrm{eq}}(\sigma_0^+)\rp^2,
\label{e8.13}\\
\lambda&\to&\frac{3\kB}{2m}\kappa+
\frac{75}{64}\lp\frac{\kB T}{\pi m}\rp^{1/2}
\frac{1}{\sigma_0\g_2^{\mathrm{eq}}(\sigma_0^+)}\lp1+
\frac25\pi n\sigma_0^3\g_2^{\mathrm{eq}}(\sigma_0^+)\rp^2.\label{e8.14}
\eea
Relations \refp{e8.12}--\refp{e8.14} are obtained from 
\refp{e8.4}--\refp{e8.6} with taking into account \refp{e8.8} are identical 
to those ones from the SET (RET) theory which are obtained by means of 
the standard Chapman-Enskog procedure.

\subsection{Square-well potential}

%
%
\bea
\ba{lclclclclclclcl}
\vt\veps_{li}&=&0,&\quad&
\vt s_i&=&0,&\quad&i&=&1,\ldots,n^*,&\qquad&\vt\veps_{r1}&\ne&0,\\
\vt\veps_{rj}&=&0,&\quad&
\vt s_j&=&0,&\quad&j&=&2,\ldots,m^*,&\qquad&\vt s_1&=&\beta\veps.\\
\ea\label{e8.15}
\eea
In this case initial MSPI \refp{e2.2} transfers into ``square-well'' one of 
the DRS (RDRS) theory. Defining $\sigma_{r1}=\sigma$, 
$\vt s_1=\vt s=\beta\vt\veps_{r1}\equiv\beta\veps$, where $\veps$ is the 
square-well depth, and taking into account \refp{e8.15} one obtains:
\bean
\Lambda&\to&\sigma_0^3\g_2^{\mathrm{eq}}(\sigma_0^+)+\sigma^3\g_2(\sigma^-)
\lc\e^{-\vts s}-1\rc,\\
\lambda^*&\to&\frac12\lc\sigma_0^2\g_2^{\mathrm{eq}}(\sigma_0^+)+
\sigma^2\g_2^{\mathrm{eq}}(\sigma^-)\e^{-\vts s}\Xi(\vt s)\rc,\\
H_1&\to&\frac{\sqrt{\pi}}{2}\Lambda+\sigma^3
\g_2^{\mathrm{eq}}(\sigma^-)\vt s
\e^{-\vts s}\lc\frac{\sqrt{\pi}}{4}-\frac13\vt s^{3/2}+
\frac13\e^{\vts s}\Gamma\lp\frac52,\vt s\rp\rc,\\
H_2&\to&\sigma_0^4
\g_2^{\mathrm{eq}}(\sigma_0^+)+\sigma^4
\g_2^{\mathrm{eq}}(\sigma^-)\e^{\vts s}\Xi(\vt s),\\
H_3&\to&\frac{\sqrt{\pi}}{2}+\e^{\vts s}\Gamma\lp\frac35,\vt s\rp,\\
D_1&\to&\sigma^3
\g_2^{\mathrm{eq}}(\sigma^-)\vt s\e^{-\vts s}H_3(\vt s),\\
D_2&\to&\sigma^2\g_2^{\mathrm{eq}}(\sigma^-)\vt s^2\e^{-\vts s},\\
\Xi(\vt s)&=&\e^{\vts s}-\frac12\vt s-2\int_0^\infty\d x\;
x^2\sqrt{x^2+\vt s},
\eean
%
%
%
%
%
\bea
a^{(1)}&\to&\frac{\frac{15}{4}n\lp1+\frac25\pi n\Lambda\rp+
\frac32\sqrt{\pi}n^2\sigma^3\g_2^{\mathrm{eq}}(\sigma^-)\vt s\e^{-\vts s}
H_3(\vt s)}{8\sqrt{2\pi}n^2\lc\lambda^*+
\frac{11}{32}\sigma^2\g_2^{\mathrm{eq}}(\sigma^-)\vt s^2\e^{-\vts s}\rc},
\label{e8.16}\\
b^{(0)}&\to&\frac{5n\lp1+\frac{4}{15}\pi n\Lambda\rp+
\frac43\sqrt{\pi}n^2\sigma^3\g_2^{\mathrm{eq}}(\sigma^-)\vt s\e^{-\vts s}
H_3(\vt s)}{8\sqrt{2\pi}n^2\lc\lambda^*+
\frac{1}{12}\sigma^2\g_2^{\mathrm{eq}}(\sigma^-)\vt s^2\e^{-\vts s}\rc}
\lp\frac{m}{2\kB T}\rp^{1/2},\nonumber
\eea
and
%
%
\bea
p&\to&n\kB T\lp1+\frac23\pi n\Lambda\rp,\label{e8.17}\\
\kappa&\to&\frac49n^2\sqrt{\pi m\kB T}H_3,\label{e8.18}\\
\eta&\to&\frac35\kappa+\frac12n\kB T\lc1+\frac8{15}\sqrt{\pi}nH_1\rc 
b^{(0)},\label{e8.19}\\
\lambda&\to&\frac{3\kB}{2m}\kappa+\frac54n\kB\lp\frac{2\kB T}{m}\rp^{1/2} 
\lc1+\frac45\sqrt{\pi}nH_1\rc a^{(1)}.\label{e8.20}
\eea
Relations for transport coefficients \refp{e8.18}--\refp{e8.20} coincide 
with those ones for DRS (RDRS) theory which were obtained by means of the 
standard Chap\-man-Ens\-kog procedure. Of course, only the first 
approximation on gradients of hydrodynamic parameters is implied everywhere.

\subsection{Smooth long-range potential}

Finally let us consider briefly the case, when
%
%
\bea
\ba{lclclclclclclcl}
\vt\veps_{li}&\to&0,&\quad&
\vt\sigma_{li}&\to&0,&\quad&i&=&1,\ldots,n^*,&\quad&n^*&\to&\infty,\\
\vt\veps_{rj}&\to&0,&\quad&
\vt\sigma_{rj}&\to&0,&\quad&j&=&1,\ldots,m^*,&\quad&m^*&\to&\infty,\\
\ea\label{e8.21}
\eea
and additional condition for \refp{e8.21}:
%
%
\bea
\ba{lcl}
-\ds\frac{\vt\veps_{li}}{\vt\sigma_{li}}&\to&\phi_t'(\sigma_{li}),
\\ [1ex]
\phantom{-}\ds
\frac{\vt\veps_{rj}}{\vt\sigma_{rj}}&\to&\phi_t'(\sigma_{rj}).\\
\ea\label{e8.22}
\eea
From the geometrical point of view, \refp{e8.21} and \refp{e8.22} 
correspond to the case when MSPI \refp{e2.2} is ``merged'' with some smooth 
long-range potential $\phi_t$ at $r>\sigma$. It can be shown that in this 
case
%
%
\bea
p\to n\kB T\lp1+\frac23\pi n\sigma_0^3
\g_2^{\mathrm{eq}}(\sigma_0^+)-\frac23\pi n^2
\int\d r\;r^3\g_2^{\mathrm{eq}}(r)\phi_t'(r)\rp.\label{e8.23}
\eea
Expressions for $\kappa$, $\eta$ and $\lambda$ are completely similar to 
\refp{e8.12}--\refp{e8.14} of the SET (RET) theory with only difference in 
the form for $\g_2^{\mathrm{eq}}(\sigma_0^+)$. In SET (RET) 
$\g_2^{\mathrm{eq}}(\sigma_0^+)$ is some binary equilibrium correlation 
function of hard spheres on contact whereas here it is binary equilibrium 
correlation function of a system with interaction potential of type of hard 
spheres plus some long ``tail'' $\phi_t$, $r>\sigma_0$. Thus, one obtains 
final relations for $p$ \refp{e8.23} and $\kappa$, $\eta$ and $\lambda$ of 
the KMFT theory \cite{15}.

\section{Numerical calculations}

First of all, let us remember that in the theory under consideration we 
deal with some multistep potential of interaction \refp{e2.2}. If one has 
any real information about real (smooth, of course) potential of 
interaction, it should deal with large amount of definition parameters. But 
really, when interaction potential is known, the number of independent 
master parameters is shortened greatly. That is the necessary condition, 
because a model interaction potential should approximate real one more or 
less correctly. First question appearing here consists in the 
representation of an initial smooth interaction potential by a multistep 
one. Let us consider one possible way of definition in which all distances 
between walls of the same kind are equal, i.e.: $\vt\sigma_{li}=\const$, 
$i=1,\ldots,n^*$, $\vt\sigma_{rj}=\const$, $j=1,\ldots,m^*$. Then, to 
define model interaction potential one needs $\sigma_0$ -- position of hard 
sphere wall, $\sigma_{\max}$ -- position of the most removed attractive 
wall ($\sigma_{\max}=\sigma_{r_{m^*}}$), $n_p$ -- the number of short 
lengths dividing repulsive area $[\sigma_0,\sigma_{\mathrm{mean}}]$ and 
$m_p$ -- the number of short lengths dividing attractive area 
$[\sigma_{\mathrm{mean}},\sigma_{\max}]$, where $\sigma_{\mathrm{mean}}$ is 
the minimum position of a real interaction potential. Now MSPI is built. 
Numbers of repulsive $n^*$ and attractive $m^*$ walls are uniquely 
determined via numbers of dividing lengths $n_p$ and $m_p$. In this 
representation of some real interaction potential by MSPI, one realizes 
original entwining of model potential around real one.

Second question is the problem of optimal dividing, i.e. definition of 
parameters $\sigma_0$, $\sigma_{\max}$, $n_p$, $m_p$ in such a manner to 
obtain good results right in the first approximation. We tried to solve 
this problem numerically.

Numerical computations of transport coefficients were carried out for Argon 
with the Lennard-Jones potential
%
%
\bea
\phi_{\mathrm{real}}\simeq\phi_{\mathrm{LJ}}=4\veps_{\mathrm{LJ}}
\ls\lp\frac{\sigma_{\mathrm{LJ}}}{r}\rp^{12}-
\lp\frac{\sigma_{\mathrm{LJ}}}{r}\rp^6\rs,\label{e9.1}
\eea
where $\sigma_{\mathrm{LJ}}=3.405$ \AA, $\veps_{\mathrm{LJ}}/\kB=119.8$ K.

Starting point in numerical analysis of transport coefficients of our 
theory are relations \refp{e8.4}--\refp{e8.6} with additional equation for 
binary equilibrium correlation function $\g_2^{\mathrm{eq}}$ of a system 
with potential in a form of multistep function. In our calculation we used 
for $\g_2^{\mathrm{eq}}$ the following approximation:
%
%
\bea
\g_2^{\mathrm{eq}}(r)&=&\g_2^{(0)}(r)\exp\{-\beta\phi(r)\},\label{e9.2}\\
\phi(r)&\equiv&\vphi(r),\nonumber
\eea
where $\g_2^{(0)}(r)$ is the binary equilibrium correlation function of 
hard spheres of diameter $\sigma_0$. Its analytical expression is well 
known \cite{23}.
\begin{table}[ht]
\caption{Parameters for different theories and calculations for transport 
coefficient $\eta$. Bottom part contains square displacement of results of 
SET (RET), MET (BH), DRS (RDRS) theories and our theory denoted by GDRS 
(i.e. generalized DRS) from MD simulation. The GDRS result is seen to be 
the most close to MD simulation. The same parameters were used for 
calculation of another transport coefficients.}
\label{table1}
\vspace*{1ex}
\begin{center}
\small
\begin{tabular}{lllll}
\hline\hline
SET (RET)	&\multicolumn{4}{|l}{SIGMZ0=1.047}\\
MET (BH)	&\multicolumn{4}{|l}{$\sigma_0(T)=\sigma_{\mathrm{LJ}}
		\ds\frac{1.068+0.3837\lp\kB T/\veps_{\mathrm{LJ}}\rp}
		{1.000+0.4293\lp\kB T/\veps_{\mathrm{LJ}}\rp}$}\\
DRS (RDRS)	&\multicolumn{4}{|l}{SIGMZ0=0.891, SIGMZM=1.342, EZDRS=0.929}\\
GDRS		&\multicolumn{4}{|l}{SIGMZ0=0.940, SIGMZM=1.960, $n_p$=3,
		$m_p$=9, $n^*$=2, $m^*$=6}\\

\hline\\
\hline
MD		&SET (RET)	&MET (BH)	&DRS (RDRS)	&GDRS\\\hline
0.0		&0.00125	&0.00794	&0.000217	&0.000206\\
\hline\hline
\end{tabular}
\end{center}
\end{table}

\begin{figure}[ht]
\begin{center}
\fbox{\includegraphics*[bb=59 53 547 609,%
		angle=-90,width=0.95\hsize]{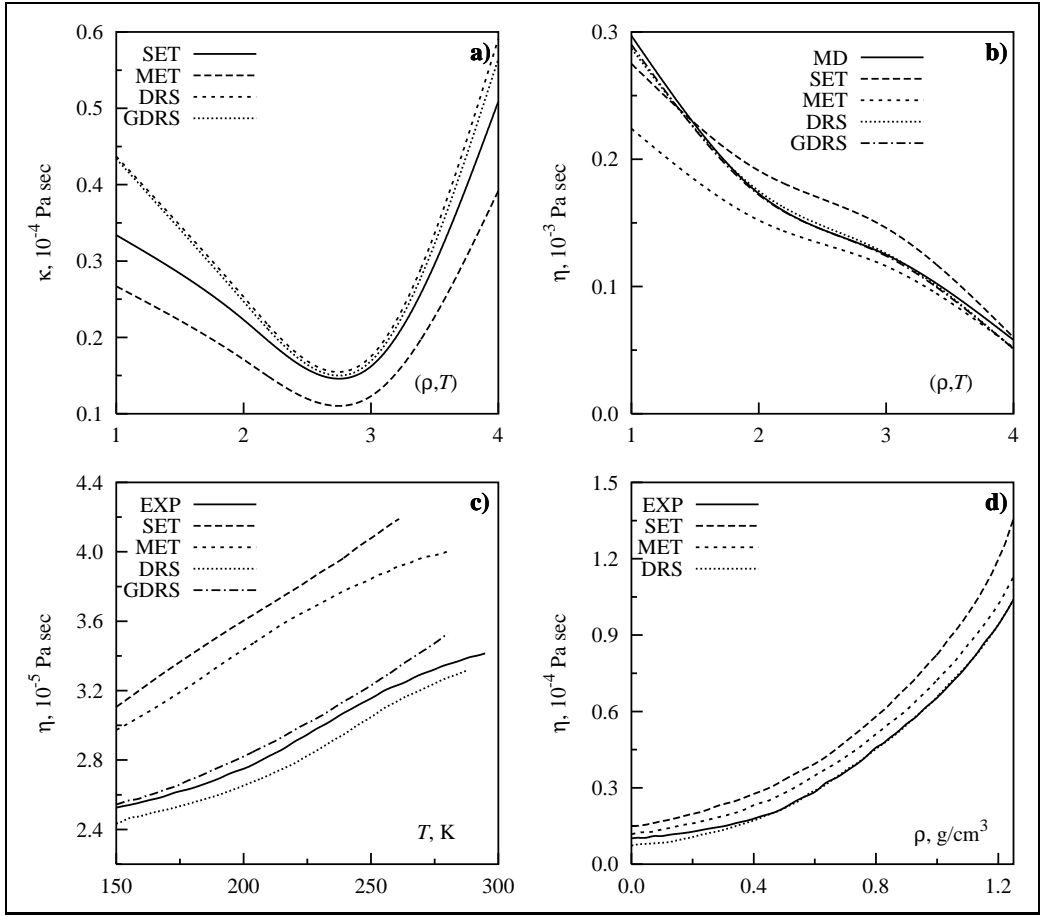}}
\end{center}
\caption{Transport coefficients for Argon. a) bulk viscosity $\kappa$ along 
the liquid-vapour curve. $x$-axis is in units of $\rho$(g/cm$^3$), 
especially: 1.4327, 1.4180, 1.1621 and 0.8017 for 1, 2, 3 and 4, 
correspondingly. b) shear viscosity $\eta$. $x$-axis is in units of 
($\rho$(g/cm$^3$),$T$(K)), especially: $\rho_1=1.43$, $T_1=83.9$, 
$\rho_2=1.28$, $T_2=104.5$, $\rho_3=1.16$, $T_3=119.56$ and $\rho_4=0.802$, 
$T_4=147.1$. c) $\eta=\eta(T)$ at $\rho=\rho_{\mathrm{cr}}$; d) 
$\eta=\eta(\rho)$ at $T=139.7$ K. Experimental data plotted in c) and d) 
are borrowed from \protect\cite{24}.}
\label{figure3}
\end{figure}

\begin{figure}[ht]
\begin{center}
\fbox{\includegraphics*[bb=59 54 547 615,%
		angle=-90,width=0.95\hsize]{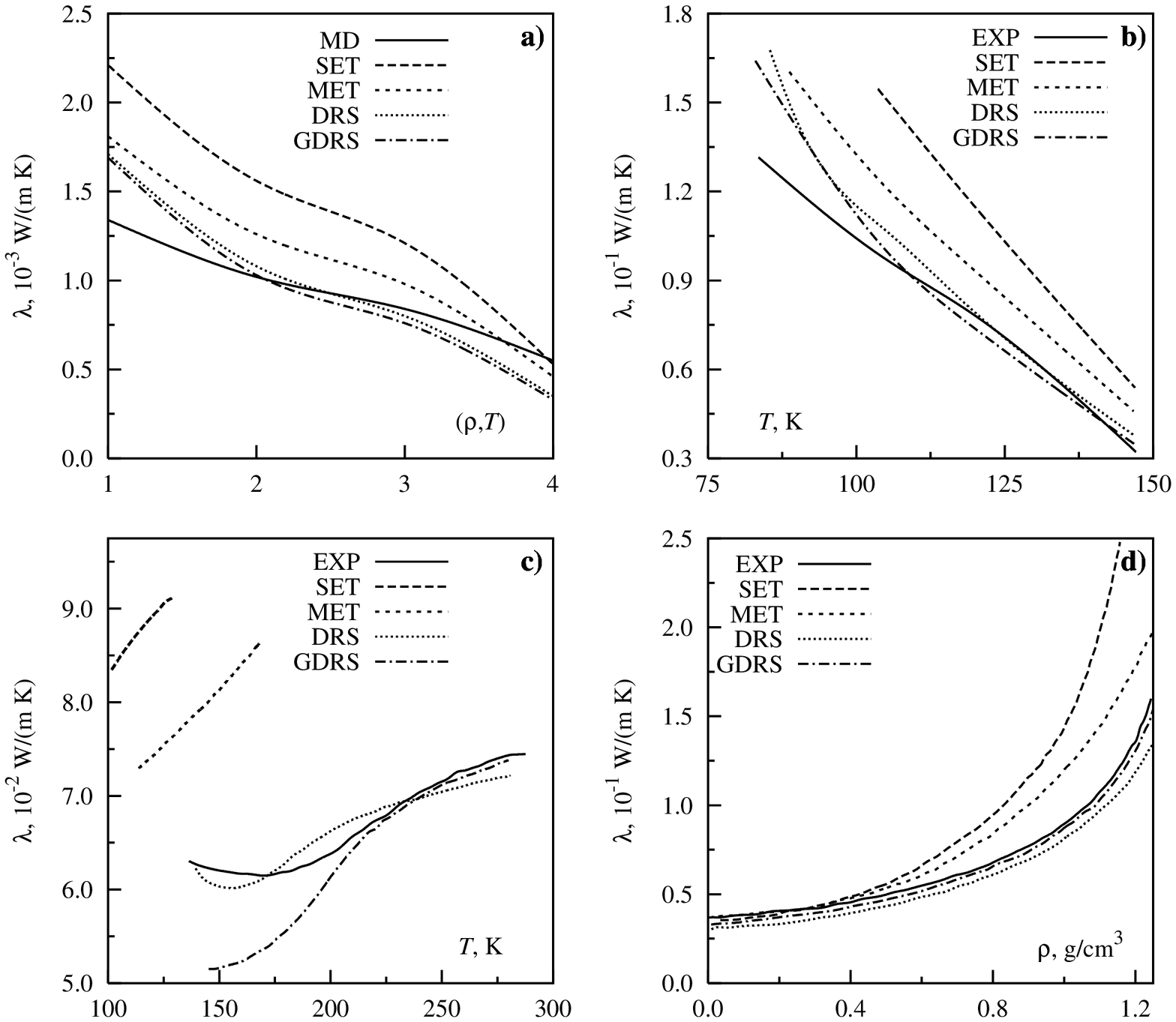}}
\end{center}
\caption{Thermal conductivity $\lambda$ of Argon. a) MD simulations and 
different theories calculations in the same points as in Fig. 3 b). b) 
$\lambda=\lambda(T)$ at $\rho=2\rho_{\mathrm{cr}}$, different theories are 
compared with experimental data. c) $\lambda=\lambda(T)$ at 
$\rho=2\rho_{\mathrm{cr}}$; d) $\lambda=\lambda(\rho)$ at $T=298$ K. All 
experimental data plotted in this figure are borrowed from 
\protect\cite{24}.}
\label{figure4}
\end{figure}

First, one calculates transport coefficients along gas-liquid saturation 
curve. There were 5 points of calculation ($\rho_i=mn_i$, $T_i$, 
$i=1,\ldots,5$) along curve of saturation for which such transport 
coefficient as shear viscosity $\eta$ is known from the MD simulation 
\cite{15}. MSPI parameters $n_p$, $m_p$, 
SIGMZ0=$\sigma_0/\sigma_{\mathrm{LJ}}$, SIGMZM=$\sigma_{\max}/\sigma_0$ 
were defined from the minimum of square displacement of the theory from 
corresponding MD results. Parameters of the DRS (RDRS) theory were defined 
in much the same way: SIGMZ0=$\sigma_0/\sigma_{\mathrm{LJ}}$, 
SIGMZM=$\sigma/\sigma_0$, EDRS=$\veps/\veps_{\mathrm{LJ}}$, as well as for 
SET (RET) theory: SIGMZ0=$\sigma_0/\sigma_{\mathrm{LJ}}$. 
Table~\ref{table1} shows the results.
\begin{table}[ht]
\caption{Transport coefficients $\kappa$, $\eta$ and $\lambda$ calculated 
by different theories.}
\label{table2}
\begin{center}
\begin{tabular}{lllll}
\hline\hline
\multicolumn{5}{c}{Bulk viscosity $\kappa$, 10$^{-4}$ Pa sec}\\
\hline
$\rho$, g/cm$^3$&SET		&MET		&DRS		&GDRS\\
\hline
1.4327		&0.33387	&0.26739	&0.43672	&0.43371\\
1.4180		&0.32270	&0.25654	&0.40946	&0.40538\\
1.2777		&0.22253	&0.17126	&0.25314	&0.24708\\
1.1621		&0.16222	&0.12277	&0.17466	&0.16928\\
0.8017		&0.05092	&0.03919	&0.05914	&0.05653\\
\hline\\
\multicolumn{5}{c}{Shear viscosity $\eta$, 10$^{-3}$ Pa sec}\\
\hline
$\rho$, g/cm$^3$&SET		&MET		&DRS		&GDRS\\
\hline
0.2970		&0.27460	&0.22428	&0.28794	&0.28953\\
0.2620		&0.26633	&0.21627	&0.27144	&0.27189\\
0.1734		&0.19113	&0.15248	&0.17491	&0.17248\\
0.1255		&0.14577	&0.11622	&0.12627	&0.12383\\
0.5790		&0.06014	&0.05210	&0.05134	&0.05087\\
\hline\\
\multicolumn{5}{c}{Thermal conductivity $\lambda$, W/(m K)}\\
\hline
$T$, K		&SET		&MET		&DRS		&GDRS\\
\hline
\n83.90		&0.22107	&0.18186	&0.17078	&0.16850\\
\n86.50		&0.21468	&0.17566	&0.16187	&0.15877\\
104.50		&0.15622	&0.12602	&0.10790	&0.10325\\
119.56		&0.12080	&0.09763	&0.08029	&0.07603\\
147.10		&0.05254	&0.04608	&0.03484	&0.03309\\
\hline\hline
\end{tabular}
\end{center}
\end{table}
Table~\ref{table2} shows all results 
of calculation of transport coefficients by different theories. Their 
comparison with experimental data and MD simulations are presented in 
Figs.~\ref{figure3} and~\ref{figure4}. It is clear to see that GDRS results 
practically coinside with experimental data in a wide range of densities 
and temperatures.

\section{Concluding remarks}

Let us discuss areas of application of kinetic equation \refp{e2.13}. We 
should remember conditions of general derivation of this equation 
within the frame of Bogolubov-Zubarev approach \cite{25,26}. Specific 
demand to the geometry of a potential and to the density of a system 
is that the mean free path $l_{\mathrm{f}}$ should be greatly less 
than $\vt\sigma$ -- a minimal clearance between the walls. In such a way, 
one should expect that bigger distance between walls and higher density 
give smaller error in the kinetic equation. This error is introduced by 
the limiting condition for interaction time $|\tau|\to+0$ \cite{26}. On 
the other hand, next to the theory error there is an error caused by a 
deviation of the multistep potential of interaction from a real one. Real 
potential is smooth and the error is less when clearance between walls is 
smaller. In the limit \refp{e8.21} this error is the smallest. One can 
observe that these two types of errors have opposite tendencies. So, to 
apply obtained kinetic equation to systems with real smooth interparticle 
interaction potential in view of a geometry of MSPI one should find a 
compromise solution. Firs of all, MSPI should approximate real potential 
more or less well. At the same time the condition 
$l_{\mathrm{f}}\ll\;\vt\sigma$ must obey. This raises the question of 
optimal dividing of a real potential of interaction on some multistep one. 
Density decreasing makes impossible obtaining the Boltzmann analogue from 
the equation under consideration in the limit $n\to0$. Let us evaluate 
numerically. Let $\sigma_0$ is the position of a hard sphere, $\vt\sigma$ 
is the minimal distance between walls, $\sigma_{\max}\simeq2\sigma_0$ is 
the location of the most removed attractive wall. It is well known from 
the theory of rarefied gases \cite{1,27} that the mean free path 
$l_{\mathrm{f}}\approx1/\sqrt{2}\pi n\sigma_{\max}^2$. In dense gases it 
decreases in the first approximation in $\g_2^{\mathrm{eq}}(\sigma_0^+)$ 
times where $\g_2^{\mathrm{eq}}(\sigma_0^+)$ is contact value of binary 
equilibrium correlation function \cite{19}. Thus, 
$l_{\mathrm{f}}\approx1/4\sqrt{2}\pi n\sigma_0^2
\g_2^{\mathrm{eq}}(\sigma_0^+)$. Introducing dimensionless density 
$\Delta=\frac16\pi n\sigma_0^3$, one obtains:
%
%
\bea
\frac{\vt\sigma}{\sigma_0}\gg\frac{1}{24\sqrt{2}\pi\Delta
\g_2^{\mathrm{eq}}(\sigma_0^+)}=\gamma.\label{e10.1}
\eea
For $\Delta=0.25$ and $\g_2^{\mathrm{eq}}(\sigma_0^+)\simeq2.5$ one 
obtains $\gamma\approx1/25$. As far as initial preconditions of the theory 
do not obey then in the limit \refp{e8.21}, the theory error reaches its 
maximal value. However, kinetic equation transfers then into the equation 
of the kinetic mean field theory \cite{15}.

\end{document}